\documentclass[manuscript]{aastex}
\usepackage{psfig}
\usepackage{emulateapj5}
\normalsize

\newcommand{\mjyb}{mJy beam${}^{-1}$}
\begin{document}

\title{A Wide Field, Low Frequency Radio Image of the Field of M31: 
II. -- Source Classification and Discussion} 
\author{Joseph D. Gelfand}
\affil{Harvard-Smithsonian Center for Astrophysics}
\affil{60 Garden St. MS-10 Cambridge, MA 02138}
\email{jgelfand@cfa.harvard.edu}
\and
\author{T. Joseph W. Lazio}
\affil{Naval Research Laboratory - Code 7213}
\affil{4555 Overlook Ave. SW Washington, DC 20375-5351}
\email{Joseph.Lazio@nrl.navy.mil}
\and
\author{B. M. Gaensler}
\affil{Harvard-Smithsonian Center for Astrophysics}
\affil{60 Garden St. Cambridge, MA 02138}
\email{bgaensler@cfa.harvard.edu}

\begin{abstract}
We have previously presented the results of a 325~MHz radio survey of M31, 
conducted with the A-configuration of the Very Large Array.  In this survey, a 
total of 405 radio sources between $\la$6\arcsec~ and 170\arcsec~in extent 
were mapped with a resolution of 6\arcsec~and a sensitivity of 
$\sim$0.6~mJy~beam${}^{-1}$.  Here, we compare the resultant source list and 
image with other radio, IR, optical, X--ray observations and catalogs of the 
region.  Through this, we were able to identify five supernova remnant (SNR) 
candidates and three pulsar wind nebula (PWN) candidates in M31, as well as 
three Milky Way radio stars, a possible Milky Way Planetary Nebula, and a bevy 
of interesting extragalactic objects: a BL Lac, a Giant Radio Galaxy, a galaxy 
merger, and several high-z radio galaxy candidates.  In addition, a large 
number of compact ($\theta \la 6\arcsec$) extremely steep spectrum sources 
($\alpha \leq -1.6; S_{\nu} \propto \nu^{\alpha}$) were detected whose nature 
is unknown.
\end{abstract}

\section{Introduction}
\label{intro}
The radio population of a star-forming galaxy -- dominated by supernova 
remnants (SNRs), {\sc Hii} regions, and radio pulsars -- contains much 
information on the evolution and properties of such galaxies.  To better 
understand such sources in M31, we have surveyed M31 at 325 MHz with the Very 
Large Array\footnote{The VLA is operated by The National Radio Astronomy 
Observatory, which is a facility of the National Science Foundation operated 
under cooperative agreement by Associated Universities, Inc.} (VLA) in order 
to identify radio sources intrinsic to this galaxy.  In the introduction of 
\citeauthor{paper1} \citeyear{paper1} (hereafter referred to as Paper I), we 
discussed some of the advantages of surveying M31 at low frequencies, mainly 
 that the large field of view of the VLA at this frequency allows one to image 
the optical disk of M31 in one pointing.  However, observing at low frequencies
 also has advantages in classifying the detected sources.  Low frequency 
observations ($\nu \la 1$~GHz) have increased sensitivity to the 
steep--spectrum ($\alpha \la -0.5$) population of a galaxy, mostly composed of 
SNRs and radio pulsars, than higher frequency observations.  These two 
different class of objects provide separate pieces of information: the SNR 
population traces recent massive star formation and individual SNRs both 
highlight the local interstellar medium (ISM) and are crucial to understanding 
the cosmic ray population of the galaxy, while radio pulsars are very useful 
in determining the line-of-sight electron density and magnetic field and 
contain information on the total core-collapse supernova rate of a galaxy.  
In a low-frequency survey, these sources are noticeably brighter than the 
flat--spectrum population of a galaxy, composed mainly of HII regions and 
Pulsar Wind Nebulae (PWNe).  By comparing our images to higher frequency 
images of M31, we can separate these different components of the radio 
population of M31 from each other.

In addition, low--frequency observations coupled with higher--frequency ones 
allow the detection of spectral curvature, a powerful probe 
of a source and its surroundings.  Two processes which can cause 
low--frequency spectral turnover are free--free absorption, indicative of 
ionized material along the line of sight, and synchrotron self--absorption, 
which is believed to occur in active galactic nuclei (AGN; \citeauthor{odea98}
 \citeyear{odea98}).  Simultaneous 74 and 333 MHz 
observations have detected free-free absorption caused by ionized gas inside 
SNR Cas A \citep{casa}, and 74 and 327 MHz observations of SNR W49B discovered 
a foreground HII region obscured by a molecular cloud \citep{w49b}.  
Low--frequency surveys of the center of star-burst galaxy M82 allowed the 
authors to calculate the properties of its ionized ISM through the detection 
of free-free absorption (e.g. \citeauthor{wills} \citeyear{wills}; 
\citeauthor{allen} \citeyear{allen}).  In this case, unlike for the
the Milky Way cases discussed above, spectral turnover was detected
above 325~MHz.  Low frequency 
spectral turnover has also been observed in compact, distant, powerful radio 
galaxies known as Gigahertz Peaked sources (GPS) believed to be examples of 
AGN powered radio galaxies whose jets have yet to penetrate through the ISM of
the galaxy \citep{odea98}.  In addition, a radio spectrum with low--frequency 
values {\it above} the power--law extrapolation of higher frequency 
measurements is the signature of nonlinear shock acceleration, an important 
process in explaining the observed cosmic ray spectrum, and has been observed 
in young SNRs such as Tycho's SNR and Kepler's SNR \citep{ellison}.

In Paper I, we described a 325 MHz (``GLG'') survey of M31 conducted
with the A-configuration of the VLA.  In this survey, we imaged 405
radio sources between 6\arcsec~(the beam size)~and 170\arcsec~in
extent, and calculated for each source its angular distance from the
center of M31 ($\theta_{M31}$), the projected radius of the source
assuming it is in the optical disk of M31 ($R_{M31}$), its major axis
($\theta_M$), minor axis ($\theta_m$), position angle ($\theta_{PA}$), peak
intensity ($A$), integrated flux ($S$), and - through comparisons with
the 36W \citep{36w}, 37W \citep{37w}, Braun \citep{braun}, and NVSS
\citep{nvss} radio surveys - its spectral index $\alpha$ and spectral
curvature parameter $\varphi$, using the convention:
\begin{equation}
S_{\nu} \propto \nu^{\alpha} e^{-\varphi(\nu^{-2.1})}
\end{equation}
To determine if a ``GLG'' source had a counterpart in one of these other 
catalogs, we used a limiting distance of:
\begin{equation}
\label{radctrp}
r=Max\left(\frac{\theta_M^{GLG}+\theta_m^{GLG}}{2},\frac{\theta_M+\theta_m}{2}\right)
\end{equation}
where $\theta_M^{GLG}$ is the major axis of the GLG source, $\theta_m^{GLG}$ 
is the minor axis of the GLG source (these are not the deconvolved values but 
the size of the source in our images), $\theta_M$ is the deconvolved major 
axis of the catalog source, and $\theta_m$ is the deconvolved minor axis of 
the catalog source \citep{braun}.  Only a ``good match'' -- one where the 
GLG source had only one match in the other catalog and its counterpart only 
had no other matches in the GLG source list -- was used to calculate 
$\alpha$ and $\varphi$.

In addition, we classified each source into one of four morphological 
categories:
\begin{itemize}
\item Unresolved (``U''): The 281 GLG sources with 
$\theta_M < 2 \times \theta_m$ and $\theta_M \la 15$\arcsec.
\item Elongated (``El''): The 16 sources with $\theta_M > 2 \times \theta_m$ 
and $\theta_M \la 15$\arcsec.
\item Complex (``C''): The 51 sources with $\theta_M \ga 15$\arcsec~that have 
structure on scales smaller than $\sim$10\arcsec.
\item Extended (``Ex''): The 57 sources with $\theta_M \ga 15$\arcsec~that do 
not have structure on scales smaller than $\sim$10\arcsec.  These sources were 
detected by running the source-finding algorithm {\sc SFIND}, part of the 
{\sc MIRIAD} software package, on images smoothed to 
20\arcsec$\times$20\arcsec.  A subset of these category are the Good Extended 
Sources (``ExG''), the two sources which satisfy the above criteria and have a 
counterpart in one of the aforementioned higher frequency radio surveys.  
Extended sources without counterparts in one of these surveys are {\it most 
likely spurious}, but were included in the final source list for completeness.
\end{itemize}
This source list is provided in Table 3 of Paper I.

As described in Paper I, a statistical analysis of the angular distribution 
($\theta_{M31}$) of sources, the radial distribution of sources ($R_{M31}$), 
flux density ($S_{325}$), and spectral index ($\alpha$) distribution of 
sources revealed that the majority of detected sources are background radio 
galaxies.  The primary purpose of this paper is to identify sources intrinsic 
to M31, determine their nature, and discuss their properties.  In addition, we 
have also identified interesting background and foreground sources.  This 
paper is structured as follows: Section \ref{classification} discusses the 
information used to classify sources, Section \ref{interesting} presents the 
final classification of sources, and Section \ref{conclusion} summarizes 
our results.  

In this paper, $S_{\nu}$ is the flux density of a source at a 
frequency $\nu$ MHz, and the distance to M31 is assumed to be 780$\pm$13~kpc 
\citep{stanek}.

\section[]{Classification of Observed sources}
\label{classification}

In Paper I, we provided statistical information which helped determine the 
general properties of the GLG sources.  However, we also want to determine the 
nature of each observed source in order to sort out background radio galaxies 
from sources in M31 and the Milky Way, as well as to differentiate between the 
different kinds of sources that comprise these categories.  In order to 
accomplish this, we used three different pieces of information: a source's 
radio spectral properties, its morphological characteristics, and its 
multi-wavelength (from the far--IR to the X--ray) properties.  While it was 
not possible to determine the nature of all 405 GLG sources, we were able to 
classify 125 of them based on the information and methodology described below.

\subsection{Radio Spectrum Classifications}
\label{specclass}
We defined three spectral categories in order to help separate different 
classes of sources -- steep spectrum ($\alpha \leq -1.3$~and 
$\mid\varphi\mid \leq 3\sigma_{\varphi}$, where $\sigma_{\varphi}$ is
 the error in $\varphi$), flat spectrum ($-0.5 \leq \alpha \leq 0.5$~and~
$\mid\varphi\mid \leq 3\sigma_{\varphi}$), and sources with spectral curvature 
( $\mid\varphi\mid > 3\sigma_{\varphi}$).\footnote{Only ``U'', ``El'', ``C'', 
and ``ExG'' sources are included in these categories, since the 55 ``Ex'' are 
unlikely to be real sources.  See Section 2.3 of Paper I for more
details.}  In the cases of sources with spectral curvature, the source
 must have been detected at two other radio frequencies -- most 
commonly 610 and 1400~MHz.  105 GLG sources fell into at least one of
these categories.  Table \ref{steepspec} lists the steep spectrum
sources, Table \ref{flatspec} lists the flat spectrum sources, and
Table \ref{turnspec} lists the sources with spectral curvature.
Figure \ref{specsrcs} shows the positions of these sources with
respect to the optical disk of M31.  The distribution of steep
spectrum and flat spectrum sources is uniform on the sky, while the
distribution of turnover sources is condensed around the optical disk
of M31.  However, this most likely is a selection effect -- as
mentioned in Paper I, $\varphi$ was only calculated for those GLG
sources with counterparts in at least three other radio surveys.  As
shown in Figure 1 of Paper I, these surveys only overlap around the
optical disk of M31.

This classification scheme was defined in such a way to separate different 
physical classes of sources.  

\paragraph{Steep Spectrum Sources}
With the advent of deep low--frequency surveys, Ultra-Steep Spectrum sources 
(USS;~$\alpha \leq -1.3$) have been increasingly studied and were discovered 
to be efficient tracers of High-z Radio Galaxies (\citeauthor{pedani03} 
\citeyear{pedani03}; \citeauthor{hzrg} \citeyear{hzrg}) and relic radio 
galaxies (\citeauthor{komis94} \citeyear{komis94}; \citeauthor{kaplan00} 
\citeyear{kaplan00}).  In addition, pulsars -- either in M31 or in
 the Milky Way -- would fall in this category, since pulsars have an average 
spectral index of $\alpha\sim-1.8$ \citep{lorimer95}.  Thirty-six 
non-Extended sources fall into the steep--spectrum category, $\sim$10\% of the 
total non-''Ex'' population.  This percentage of steep--spectrum sources is 
much higher than in previous surveys, e.g. 0.5\% in \citet{hzrg} and 
2.7\% in \citet{aaron74}.  This difference is probably due to the increased 
sensitivity of both the present survey and the higher frequency surveys used 
for comparison than those used in these previous studies.

\paragraph{Flat Spectrum Sources}
Flat spectrum sources are most likely dominated by a combination of FRI/FRII 
radio galaxies -- both are radio galaxies whose radio emission is powered by a 
central AGN, but the radio emission of an FRI is core-dominated and in general 
is less luminous that an FRII radio galaxy, whose radio emission is 
lobe--dominated \citep{fr} -- and more exotic extragalactic objects 
such as BL Lacs.  HII regions, which often have a spectral index in this 
frequency range of $\alpha \geq -0.1$, SNRs, whose spectral indices in general 
fall between $\alpha \approx -1$ and $\alpha \approx -0.3$ 
(\citeauthor{filipovic98} \citeyear{filipovic98}; \citeauthor{greensnrcat}
 \citeyear{greensnrcat}), and PWNe, which have a radio spectral index of 
$\alpha \approx -0.1$ \citep{helfand87}, in M31 or the Milky Way, would also 
fall in this category.

\paragraph{Sources with Spectral Curvature}
Sources with spectral curvature could be time-variable flat--spectrum radio 
galaxies\footnote{If the brightness of a radio source substantially changed 
between the higher frequency radio surveys and our 325~MHz observation, it 
would manifest itself as curvature in the composite radio spectrum.}, or if 
$\varphi>0$, normal galaxies (i.e galaxies whose emission is not dominated by 
a central AGN; \citeauthor{im90} \citeyear{im90}), as well as AGN, where 
spectral turnover is most likely due to synchrotron self--absorption 
\citep{odea98}.  In addition, if located in the optical disk of M31, 
$\varphi>0$ could indicate a foreground HII region along the line of sight, or 
the result of internal free-free absorption in a HII region or SNR.  An 
extragalactic source with $\varphi<0$, in addition to possibly being variable, 
could indicate the presence of a compact steep spectrum core which dominates 
the emission at low frequencies (\citeauthor{fanti90} \citeyear{fanti90}; 
\citeauthor{athreya97} \citeyear{athreya97}).  If the source is located in 
optical disk of M31, $\varphi<0$ may be the result of a HII/SNR complex -- 
if a SNR and HII region fall along the same line of sight, the HII region 
could dominate at higher-frequencies and the SNR, due to its steeper radio 
spectrum, would dominate at lower-frequencies, or a diffuse particle 
acceleration, which has been observed in young SNRs \citep{ellison}.\\
\smallskip\\
The different kinds of sources in each of these spectral 
classifications can be differentiated from each other by their morphological 
and multi-wavelength properties discussed in Sections \ref{morphclass} and 
\ref{multiwave}.  The criteria used to differentiate them is discussed in 
Section \ref{interesting}.

\subsection{Morphological Classifications}
\label{morphclass}
Classifying a source based on its morphology is the least ambiguous method  
for determining the nature of the source used in this paper.  Of the 405 
sources detected, only the 51 ``Complex'' sources, shown in Figure 
\ref{complex}, were sufficiently resolved to classify morphologically.  All 
but a few have the radio morphology of a FRI, of a FRII, or of a radio jet.  
The classification of these sources is given in Section \ref{interesting}, and 
sources with anomalous morphologies are discussed further in Section 
\ref{weirdmorph}.

\subsection{Multi-wavelength Classifications}
\label{multiwave}
In order to classify sources using their properties at other wavelengths, the
 GLG source list was compared to source lists derived from targeted near 
infrared (IR), far infrared (FIR), optical, and X-ray observations of M31, as 
well as all-sky surveys in these wavebands.  We looked for counterparts across 
the electromagnetic spectrum because the different wavebands are sensitive to 
different kinds of objects, many of which could appear in the GLG source list.
The result of these comparisons is summarized in Table \ref{mwcomp}.  In this
table, a counterpart is where a GLG source had at least one counterpart in 
this catalog and a good counterpart is the where the GLG source had only one 
counterpart in this catalog and no other GLG source had this catalog source as 
a counterpart.  For ``C'' sources, which were resolved in our survey,
a good counterpart could be spurious due to the large search radius
 used. For these sources, we checked if the counterpart corresponded 
to a region of high radio emission. The number of false counterparts
 and false good counterparts were calculated by shifting the RA and 
DEC of every source in the catalog and then determining the number of 
counterparts and ``good'' counterparts.  In Table \ref{mwcomp}, the 
number of false counterparts and false ``good'' counterparts given is 
the average number of counterparts/''good'' counterpart in the 
different shifts, and the error is the standard deviation.  The 
amount of the shift was typically $\pm$1\arcmin in both RA and DEC, 
but was larger for some catalogs (which we note when we discuss them).

\subsubsection{Near-Infrared}
\label{ir}
In order to determine if a non-extended GLG source is an IR source, we 
searched for counterparts in the 2MASS Extended Source Catalog (ESC; 
\citeauthor{2MASSesc} \citeyear{2MASSesc})\footnote{Only ESC sources within 
1$^{\circ}$ of the center of  M31 were included in this search.} and the 2MASS 
Point Source Catalog (PSC; \citeauthor{2MASSpsc} \citeyear{2MASSpsc}).  We did 
not look for IR counterparts of ``Ex'' sources because their large size, 
coupled with the high density of 2MASS sources in the direction of M31, 
resulted in a high coincidence rate, and because we believe that most ``Ex'' 
sources are spurious.  We did include ``ExG'' sources in this analysis.  For 
the ESC, the limiting distance for a counterpart between a PSC source and an 
``U'' or ``El'' GLG source is defined to be:
\begin{eqnarray} 
r=Max(3\times\sqrt{(\sigma_{\alpha}^{GLG})^2+(\sigma_{\delta}^{GLG})^2}~,
\nonumber\\
3\times\sqrt{(\theta_M^{ESC})^2+(\theta_m^{ESC})^2})
\end{eqnarray}
where $\sigma_{\alpha}^{GLG}$ is the error in RA of the GLG source, 
$\sigma_{\delta}^{GLG}$ is the error in DEC of the GLG source, 
$\theta_M^{ESC}$ is the major axis of the error ellipse of the ESC source, 
and $\theta_m^{ESC}$ is the minor axis of the error ellipse of the ESC 
source.  The errors on position were used for the ``U'' and ``El'' GLG sources 
instead of its extent because their extents are not well known.  For a complex 
GLG source, the limiting distance for a ESC counterpart was:
\begin{equation} 
r=Max\left(\frac{\theta_M^{GLG}+\theta_m^{GLG}}{2},3 \times 
\sqrt{(\theta_M^{ESC})^2+(\theta_m^{ESC})^2}\right)
\end{equation} 
where $\theta_M^{GLG}$ is the major axis of the GLG source, $\theta_m^{GLG}$ 
is the minor axis of the GLG source, and $\theta_M^{ESC}$ and $\theta_m^{ESC}$ 
are the same as before.  We used the measured size of the ``C'' GLG sources 
instead of the error on their positions because they are resolved and the IR 
emission need not be centered on the radio emission.  Once this list
was compiled, we rejected associated we considered to be unlikely due
to the large offset between the radio and ESC source.  The remaining
candidates are listed in Table \ref{jhkesc}, along with the IR
properties of the ESC counterparts.  The number of likely associations
is consistent with that expected from the analysis presented in Table
\ref{mwcomp}.

For the 2MASS PSC, counterparts were detected using the same method described 
above.  Again, we rejected associations where there was a large offset
between the radio and IR source and, for resolved radio sources, no
correspondence between the IR and radio emission.  As seen in Table
\ref{mwcomp}, the number of false good counterparts for ``U'' or
``El'' GLG sources is small.  The results of this comparison, and the
IR properties of the PSC counterparts, are listed in Table \ref{jhkpsc}.

The presence of an IR counterpart in the PSC allowed us to determine if a GLG 
source was either an FRI or FRII, background ``normal'' galaxy, a star in the 
Milky Way, or none of the above.  To determine if the observed IR magnitude 
was consistent with that of a background FRI or FRII, we first calculated the 
redshift of the GLG source if it is an FRI or an FRII by 
simultaneously solving the following equations -- the luminosity-redshift 
($L$--$z$) relation for FRI and FRIIs\footnote{This equation was 
derived using galaxies with a redshift z$<$0.5.  This analysis assumes that the
 luminosity-redshift relationship of FRIs and FRIIs does not change at higher 
redshift.}~\citep{zirbel}:
\begin{equation}
\log~L_{408}^{FRI}=(1.03\pm0.22)~\log~z + (1.92\pm0.13)
\end{equation}
\begin{equation}
\log~L_{408}^{FRII}=(26.97\pm0.32)~\log~z + (28.38\pm0.12)
\end{equation}
where $L_{408}^{FRI}$ is the luminosity in Watts 
of an FR~I galaxy at $\nu=408$~MHz, $L_{408}^{FR~II}$ is the luminosity 
in Watts of an FRII galaxy at $\nu=408$~MHz, and $z$ is the redshift of the 
FR~I/FR~II; the luminosity--distance relationship:
\begin{equation}
L_{408}=S_{408}\times 4\pi d_L^2 
\end{equation}
where $S_{408}$ is determined by using the spectral model defined in Equation 
4 in Paper I; and $d_{L}$ the luminosity distance, is defined as:
\begin{equation}
d_L=\frac{c}{H_{0}}[z+\frac{1}{2}(1-q_0)z^2]
\end{equation}
where $c$ is the speed of light, $H_0$ is Hubble's Constant (assumed to be 
71~km~sec$^{-1}$~Mpc$^{-1}$), and $q_0$, the deceleration parameter, is -0.56, 
as determined using the latest cosmological results -- for the redshift of the
 galaxy.  Using this redshift, we calculated the expected K--magnitude 
of such an FRI/FRII using the {\it K-z} relationship for radio galaxies in the 
7C Redshift survey \citep{willott}:
\begin{equation}
K=17.37+4.53~\log~z-0.31(\log~z)^2
\end{equation}
As a result of this analysis, we concluded that none of these sources
satisfied the requirement for an FRI or FRII radio galaxy since the
required redshift were ridiculously high ($z\ga10$).

We used the IR colors of PSC counterparts to determine if a GLG source is a 
background normal galaxy, a star in M31, or a star in the Milky Way. Since 
these classes of objects have very different IR colors, once we determine the 
IR colors of these objects in the field of M31 we can classify the nature of a 
GLG source based on the IR colors of its PSC counterpart.  The IR colors of 
background normal galaxies was determined by looking at the IR colors of all 
2MASS ESC sources within 1$^{\circ}$ of M31 (top-left panel of Figure 
\ref{ircolors}), since that catalog is believed to be composed almost 
entirely ($\sim$97\%) of galaxies\footnote{From ``Explanatory Supplement to 
the 2MASS All Sky Data Release'' 
(http://www.ipac.caltech.edu/2mass/releases/allsky/doc/explsup.html)}.  To 
distinguish between IR sources in the Milky Way and IR sources in M31,
 we looked at the IR color diagram of PSC sources within 5\arcmin~of the center
 of M31 -- expected to be dominated by sources in M31 (bottom-left panel of
 Figure \ref{ircolors}), and that of PSC sources $>$90\arcmin~from the 
center of M31 -- which is dominated by foreground stars and background 
galaxies (bottom-right panel of Figure \ref{ircolors}).  As Figure 
\ref{ircolors} shows, these two groups have a very different range of IR 
colors.  The band of sources in the IR color diagram of PSC sources 
$>$90\arcmin~is the same as that seen in IR color diagrams of stars in the 
Milky Way \citep{irstars}, and PSC counterparts which fall in this range are 
believe to be stars in the Milky Way.  The results of this analysis are 
presented in Section \ref{interesting}.

\subsubsection{Far-Infrared}
\label{fir}
We looked for  Far Infrared (FIR) counterpart of GLG sources in the 
Schmidtobreick et. al., 2000 (``Knots'') catalog of FIR knots in M31 
\citep{knots}, the Xu \& Helou catalog of FIR point sources in M31 \citep{xu}, 
and the IRAS Point Source Catalog (IRAS-PSC; \citeauthor{iraspsc} 
\citeyear{iraspsc}), IRAS Serendipitous Source Catalog (IRAS-SSC; 
\citeauthor{irasssc} \citeyear{irasssc}), and IRAS Faint Source Catalog 
(IRAS-FSC; \citeauthor{irasfsc} \citeyear{irasfsc}).  For these catalogs, 
matches were determined using the same criteria used for radio catalogs, 
given in Equation \ref{radctrp}.  As seen in Table \ref{mwcomp}, only
for the IRAS-PSC and IRAS-FSC catalogs are the number of counterparts
statistically significant.\footnote{The number of false counterparts 
and false good counterparts for the FIR surveys were calculated using
position shifts of 10\arcmin, instead of 1\arcmin~used for the other
surveys, because of the large reported beam size.}  However, even in
these cases the offsets are rather large.  Therefore, we are not able
to claim any overlap between the observed radio and FIR sources in
this field.  However, {\it Spitzer} with its improved angular
resolution will improve this situation.

\subsubsection{Optical}
\label{optical}
The GLG source list was compared to several optical catalogs in order to 
determine the nature of the source: the Winkler \& Williams SNR catalog 
(WW; Ben Williams, private communications 2002), the Magnier et al. SNR 
catalog (Mag; \citeauthor{magniersnr} \citeyear{magniersnr}), the Braun \& 
Walterbos SNR catalog (BW~SNR; \citeauthor{bwsnr} \citeyear{bwsnr}), the 
Walterbos \& Braun catalog of Ionized Nebulae (Ion; \citeauthor{wbion} 
\citeyear{wbion}), the Ford \& Jacoby catalog of Planetary Nebulae (FJ~PN; 
\citeauthor{fjpn} \citeyear{fjpn}), and the Pellet et. al. catalog of HII 
regions (Pellet; \citeauthor{pellet} \citeyear{pellet}).  For all of
these, we used the same criteria that we used for other radio catalogs
(Equation \ref{radctrp}).  As seen in Table \ref{mwcomp}, the
counterparts to GLG sources in the WW SNR catalog are likely to be
real but the rest are most likely coincidental. Table \ref{optctrp}
lists the counterparts of GLG sources in this catalog.

In addition to these ``specialty'' optical catalogs, we compared the GLG 
source list with the Magnier et. al. BVRI catalog of optical sources in the 
field of M31 (MagOpt; \citeauthor{magstars} \citeyear{magstars}). While a 
majority of the sources in this catalog are stars in M31, it also contains 
background galaxies and foreground (Milky Way) stars. The criteria for a 
counterpart in this catalog is the same as that used for the 2MASS PSC, 
described in Section \ref{ir}.  MagOpt counterparts of complex GLG sources 
were also determined the same way they were for the 2MASS PSC, as described in 
Section \ref{ir}.  From the analysis present in Table \ref{mwcomp}, a
majority of the counterparts between MagOpt sources and ``U'' or
``El'' GLG sources are coincidental, but due to the small offsets
between our radio sources and these sources, as well as the lack of
optical color information for most of these counterparts, it is
difficult to separate the real from the false counterparts.  As a
result, for completeness we include Table \ref{magstarctrp}, which
lists the GLG sources with a counterpart in this catalog, as well as
B, V, R, and I magnitude of the optical counterpart.

We also looked for optical counterparts of GLG sources in unpublished 
H$\alpha$, [SII], and R--band images of M31 (generously provided by Ben 
Williams).\footnote{Both the H$\alpha$ and [SII] images are 
continuum subtracted.}  These images have a larger FOV than the MagOpt catalog,
 and the H$\alpha$ \& [SII] information helps determine if a source is a SNR.
  Figure \ref{redgd} shows all GLG sources with R--band counterparts, Figure 
\ref{halpgd} shows all sources with H$\alpha$ counterparts and Figure 
\ref{siigd} shows all sources with [SII] counterparts. In the case of
GLG203, we believe the observed H$\alpha$ emission is an artifact of
the continuum subtraction process.

The final classification of GLG sources with optical counterparts is given in 
Section \ref{interesting}. 

\subsubsection{X-ray}
\label{xray}
In order to determine if a GLG source has an X-ray counterpart, we compared 
the GLG source list to source lists derived from the {\it ROSAT} All--Sky 
Survey (``RASS'') [both the faint \citep{RASSfaint} and bright 
\citep{RASSbright} 
source lists], the Second {\it ROSAT} PSPC survey of M31 \citep{rosat}, a 
{\it Chandra} survey of bright X-ray sources in M31 Globular Clusters 
\citep{m31glob}, a {\it Chandra} ACIS-I survey of the central region of M31 
\citep{acis}, a {\it Chandra} HRC survey over most of the optical disk of M31 
\citep{hrc}, a deep {\it Chandra} HRC survey of the central region of M31 
\citep{kaaret}, and a {\it XMM--Newton} survey of the northern--half of M31 
(Sergey Trudolyubov, private communication).  The method of finding 
counterparts in these X--ray catalogs was the same as that used to find 
counterparts in the 2MASS PSC survey, described in Section 
\ref{ir}.\footnote{For the ROSAT~PSPC survey, the 
size of the X--ray source used was the $\sigma_{Pos}$ (Column~9) value of the 
source given in Table 6 of \citet{rosat}.  For the Chandra~ACIS survey, this 
value is 1\arcsec~\citep{acis}, for the HRC~survey of the optical disk it is 
10\arcsec~\citep{hrc}, for the HRC~survey of the center of M31 it is 1\arcsec~
\citep{kaaret}, for the list of {\it Chandra} sources in Globular cluster it is
 5\arcsec~\citep{m31glob}, and for the {\it XMM--Newton} source list it is 
3\arcsec.}  From the analysis presented in Table 
\ref{mwcomp}, a majority of RASS counterparts and all the counterparts from HRC
 survey of the optical disk of M31 are probably coincidental.  For the Second 
{\it ROSAT} PSPC survey of M31 about one-third of the counterparts are 
coincidental, but for the {\it Chandra} ACIS-I survey of the central region of 
M31 and the {\it XMM--Newton} survey the counterparts are probably
real.  Again, for these surveys we attempted to separate the real
counterparts from the false ones by using the offsets and -- for the
resolved radio sources -- the locations of the X-ray emission with
respect to the radio emission.  Table \ref{xraygdctrp} lists the good 
X-ray counterparts of GLG sources, as well as their Hardness Ratios
(HR1 and HR2) -- the X-ray equivalent of IR colors -- where
available. HR1 and HR2 are defined as:
\begin{equation}
HR1=\frac{H1+H2-S}{H1+H2+S}
\end{equation}
\begin{equation}
HR2=\frac{H2-H1}{H2+H1}
\end{equation}
with $S$ the count rate of photons with energy $E_{\gamma}$ between 0.1 and 
0.4 KeV, $H1$ the count rate of photons with $E_{\gamma}=$0.4--0.9 KeV, and 
$H2$ the count rate of photons with $E_{\gamma}=$0.9--2.0 KeV.

In addition to cross referencing the GLG source list with the aforementioned 
X--ray source lists, we also looked for counterparts in images produced by the 
{\it XMM--Newton} survey of the optical disk of M31 (image courtesy of Albert 
Kong) and the {\it Chandra}--HRC survey of the southern disk of M31 (image 
courtesy of Ben Williams).  GLG025, GLG040, and GLG243 were discovered to have 
faint uncatalogued counterparts in the {\it XMM--Newton} image of M31,
while no new counterparts were found in the {\it Chandra}--HRC image.

A diverse group of sources have substantial X--ray emission, from 
extragalactic objects such as FRI/ FRII radio galaxies (e.g. 
\citeauthor{brunetti97} \citeyear{brunetti97}, \citeauthor{harris02} 
\citeyear{harris02}) and BL-Lacs \citep[e.g.][]{tagliaferri03} to galactic 
objects such as SNRs, pulsars and X--ray binaries.  In an attempt to 
differentiate between these sources, we looked at the HR1 vs. HR2 diagrams of 
these surveys/sources.  Unfortunately, the boundaries between these groups are 
ill--defined.  However, the work of \citet{filipovic98} showed that sources 
with HR2$<$0 are likely to be ``thermal'' SNRs, remnants whose X--ray emission 
is dominated by the ejecta and swept-up ISM.  Further classification of GLG 
sources with X--ray counterparts is given in Section \ref{interesting}.

\section{Discussion}
\label{interesting}
Based on the information described above, we have attempted to determine 
the nature of the radio sources detected in this observation.  We placed 
the GLG sources into five broad categories:
\begin{itemize}
\item Sources in M31, 
\item Sources in the Milky Way,
\item ``Background'' sources (sources neither in M31 nor in the Milky Way), 
\item Anomalous sources (source with spectral, morphological, and/or 
multi-wavelength characteristics that defy classification), and 
\item Sources for which there is insufficient information to classify them.
\end{itemize}
The first four categories will 
be discussed in more detail in this Section.  Based on the statistical 
analysis discussed in Paper I and on population synthesis work done by 
Carole Jackson (private communication), we expect that the fifth category is 
dominated by background radio galaxies.

\subsection{M31 sources}
\label{m31srcs}
The discrete radio population of a normal galaxy is dominated by SNRs, HII 
regions, radio pulsars, and Pulsar Wind Nebulae (PWNe).  Each of these classes 
of sources gives different and complementary pieces of information regarding 
the properties of M31 -- SNRs, pulsars, and PWNe give information regarding 
the death of past massive stars in M31 while HII regions provide information 
on the current massive star population of M31.  Detailed analysis of these 
objects can also provide information on the ISM and cosmic ray population of 
M31.

\subsubsection{Supernova Remnants, Pulsar Wind Nebulae, and Pulsars}
\label{snrs}
Type II SNRs and pulsars are both formed by the same phenomena -- the massive 
explosion (supernova) that marks the death of a massive star.  When the 
star explodes, it produces a large shock wave that propagates through the ISM, 
sweeping up material, that produces the SNR.  Most of the time, a rapidly 
spinning neutron star (pulsar) is produced, which in turn can generate a 
highly--relativistic particle wind that creates a Pulsar Wind Nebula (PWN) 
\citep{slane02}.  Table \ref{snrcan} lists the 11 GLG sources that we believe 
are SNR, PWN, or pulsar candidates.  The columns of Table \ref{snrcan} are as 
follows:
\begin{description}
\item[Column 1:] Name of source
\item[Column 2:] $R_{M31}$, as defined in Section 4.1 of Paper I, in kpc.
\item[Column 3:] Radio Spectral index $\alpha$
\item[Column 4:] Radio Spectral Curvature parameter $\varphi$
\item[Column 5:] Classification criteria, defined later in this section.  
GLG011 also satisfies Criteria \#2 for a High--Frequency Variable (defined in 
Section \ref{variables}), and GLG014 and GLG036 are also CESS sources 
(defined in Section \ref{cess}).
\item[Column 6:] Size of source if at the distance of M31, in parsecs
\item[Column 7:] Isotropic 325~MHz Luminosity, $L_{325}$, of the source 
assuming it is at the distance of M31, in units of mJy~kpc$^{2}$.
\item[Column 8:] Integrated radio luminosity of the source 
($\nu=10$~MHz--$10^5$~MHz) assuming the source is at the distance of M31.  This
 number does not take into account any detected spectral curvature, since if 
this source is a SNR or PWN any observed spectral turnover is most likely 
extrinsic to the source.
\item[Column 9:] Emission measure, $EM$, of the source in pc~cm$^{-6}$, if the 
spectral turnover is due to free-free emission.  The EM is only given for 
``spectral curvature'' sources with $\varphi > 0$ as defined in Section 
\ref{specclass}.  Only one SNR candidate is a ``spectral curvature'' source.
None of the PWN or pulsar candidates showed any spectral curvature.
\end{description}

A GLG source was classified as either a SNR, PWNe, or a pulsar if it was 
inside the optical disk of M31 ($R_{M31} \la 27$~kpc), was either a ``U'' or 
``El'' source, and had one of the following properties:
\begin{itemize} 
\item A good counterpart in one of the optical SNR catalogs described in 
Section \ref{optical}. (Criterion A in Table \ref{snrcan})
\item A good counterpart in the Pellet optical HII region catalog described 
in Section \ref{optical} and has $\alpha<$-0.2.  In the Milky Way, young SNRs 
are often observed to be near HII regions.  Since the progenitors of Type II 
SNRs are short--lived, they often occur in regions with ongoing massive star 
formation, which often have HII regions. (Criterion B in Table \ref{snrcan})
\item Is a ``spectral turnover'' source with $\varphi <$0.  A SNR could have 
$\varphi <$0 as a result of diffuse particle acceleration \citep{ellison}, or 
if there is a flat--spectrum source, most likely a HII region or PWNe, 
projected along the line of sight.  In this case, the SNR would dominate the 
emission observed at lower frequencies and the flat--spectrum source would 
dominate the emission at higher frequencies since the SNR has a steeper 
spectrum (in general, SNRs have a spectral index between 
$\alpha \approx -1.0~\mbox{and}~\alpha \approx -0.3$) than these objects.  A 
similar effect can be seen if a radio pulsar -- which have a median spectral 
index of $\alpha \approx -1.8$ \citep{lorimer95} -- lies along the sight.  
However it is unlikely that a pulsar in M31 would be bright enough to outshine 
a flat--spectrum source, even at 325~MHz. (Criterion C in Table \ref{snrcan})
\item A good X--ray counterpart with HR2$<$0, indicative of a thermal SNR 
\citep{filipovic98}. (Criterion D in Table \ref{snrcan})  A pulsar or PWN 
could also have a good X--ray counterpart with HR2$<$0, in the case the radio 
emission would be dominated by the PWN or pulsar while the X--ray emission is 
dominated by the thermal SNR.
\item A good counterpart in the ROSAT 2nd PSPC survey which has been classified
 as a ``SNR'' on the basis of previous optical work. (Criterion E in Table 
\ref{snrcan})
\end{itemize}
A source that met any of these criteria and has $-1.0<\alpha\leq-0.3$ was 
classified as a SNR, while a source that met this criteria and has
$\alpha > -0.3$ was classified as a PWN, and a source that met this
criteria with $\alpha \leq -1.0$ was classified as a pulsar candidate.
While not every SNR, PWNe, or pulsar in the GLG catalog will satisfy
these requirements, this set of criteria should eliminate most, if not
all, background sources. 

The distribution of these sources over the optical disk of M31, as shown in 
Figure \ref{snrpos}, appears to be relatively uniform.  Having identified the 
SNR, PWNe, and pulsar candidates in the GLG source list, we compared their 
properties to those of known objects of each type.

\paragraph{Supernova Remnants}
\label{snrcans}
The SNR candidates we identified in M31 are listed in Table
\ref{snrcan} and shown in Figure \ref{snrpics}.  For comparison
purpose, we calculated the 325~MHz flux densities ($S_{325}$) and
major axis ($\theta_M$) of known SNRs in the Milky Way (MW), LMC, SMC
and M33 if at the distance of M31.  For MW SNRs, we used radio
spectral information from the Green SNR catalog \citep{greensnrcat}
and distance information from Tables 1 and 2 of
\citet{case98}.\footnote{We did not use distances from Table 3 of
  \citet{case98} because these distances were calculated using the
  $\Sigma$--$D$ relationship, considered to be highly inaccurate.}
For LMC and SMC SNRs, we used data presented in  Table 1a of
\citet{berkhuijsen86}, while the properties of M33 SNRs was obtained
from Table 4 of \citet{gordon99}.  Figure \ref{snrbmajflux} shows
$S_{325}$ vs $\theta_M$ for these SNRs, as well as for the SNR
candidates in M31 identified above.   The GLG SNR candidates fall
within the locus of points defined by SNRs in these other galaxies,
implying that they are most likely not background sources.  However
the sample of known SNRs in the MW, LMC, SMC, and M33 are subject to a
myriad of selection effects, and there are GLG sources not classified
as SNR candidates that have similar $S_{325}$ and $\theta_M$.  From
Figure \ref{snrbmajflux}, the SNR candidates detected in our survey --
if they indeed are SNRs -- appear to be the brightest, and therefore
probably youngest, SNRs in M31.  These SNR candidates are brighter
than known radio SNRs in M31, e.g. \citet{lorant}, which were too
faint to be detected in our survey.  Higher resolution radio imaging
as well as deeper X--ray imaging are needed to verify that they are
SNRs and to better determine their properties.

Of the SNR candidates identified, only GLG193 shows spectral turnover.  If this
 spectral turnover is due to free-free absorption, the implied EM towards this 
source is $2.1 \times 10^5~$pc~cm$^{-6}$ -- more than order of magnitude 
higher than the EM of ionized gas observed inside SNRs, e.g. Cas~A which has 
an EM of $\sim 1.7 \times 10^4~$pc~cm$^{-6}$ \citep{casa}.  This possibly 
could be due to a foreground HII region, either in the Milky Way or in M31, 
since the observed EM is similar to that of Galactic HII regions 
\citep{shaver83}.  However, the lack of H$\alpha$ emission in this region, 
implies that any such HII region is obscured, possibly by a molecular cloud.  
The amount of absorption needed to block an HII region would require so much 
dust that M31 would not be visible if the obscuration was due to a Milky Way 
source.  Therefore, both the HII region and the obscuring molecular cloud need 
to be in M31.  However, this conclusion is extremely preliminary and more data 
is needed in order to determine the nature of the spectral turnover observed 
in this source.

\paragraph{Pulsar Wind Nebulae}
\label{pwnecans}
Pulsar Wind Nebulae (PWNe) are magnetic ``bubbles'' inflated by an
energetic wind produced by the central pulsar \citep{slane02}.  The PWN
candidates we identified in M31 are listed in Table \ref{snrcan} and
shown in Figure \ref{snrpics}.  The quintessential PWN 
in the Milky Way is the Crab Nebula, with a radio luminosity $L_R$ of 
$1.8 \times 10^{35}$~ergs~s$^{-1}$ \citep{helfand87} and is a bright emitter 
at X--ray, $\gamma$--ray, and even TeV energies.  Two of the three PWN 
candidates detected in the GLG survey, GLG011 and GLG198, have higher radio 
luminosities than the Crab Nebula -- GLG011 is $\sim70\times$~more luminous 
and GLG198 is $\sim30\times$~more luminous.  Both GLG011 and GLG068 have 
X--ray counterparts; the X--ray luminosity\footnote{Measured between 0.3 and 
7 keV.} ($L_X$) of GLG011 is $\sim2\times10^{36}$~erg~s$^{-1}$ (Albert Kong, 
private communication), and the X--ray luminosity\footnote{Measured between 
0.1 and 2 keV assuming a hydrogen column density 
$N_{H}=9\times10^{20}~\mbox{cm}^{-2}$ and photon index $\Gamma=-2$.} of GLG068 
is $\sim2.1 \times 10^{36}$~erg~s$^{-1}$ \citep{rosat}.  The X--ray luminosity 
of both of these sources is about an order of magnitude less than that of the 
Crab Nebula, $L_{X}^{Crab} \sim 2.5\times10^{37}$~erg s$^{-1}$ 
\citep{helfand87}.  Since the X--ray emission of a PWN reflects the current 
energy injection from the pulsar while the radio emission reflects the 
integrated energy input over the lifetime of the system, the ratio of $L_X$ to 
$L_R$ is believed to be indicative of the system's age -- though the intrinsic 
scatter in ${L_X}/{L_R}$ is large \citep{helfand87}.  The ${L_X}/{L_R}$ ratio 
of GLG011 and GLG068 is $\sim$1.6 and $\sim$21.6 respectively, well within the 
${L_X}/{L_R}$ range of known PWNe \citep{helfand87}.

GLG011 also is a High--Frequency Variable (see Section \ref{variables} for 
more information), and has a counterpart in the WENSS survey whose flux is 
3.5$\sigma$ lower than the flux measured here (numbers are in Table 
\ref{hfvcanlf}), but is not considered a Low--Frequency Variable using the 
criteria defined in Section \ref{variables}.  While PWNe are extremely dynamic 
sources, they are not known to intrinsically vary by such a degree.  This 
variation might be an example of an ``Extreme Scattering Event'' (ESE), in 
which extragalactic (outside the Milky Way) radio sources are observed to vary 
greatly in amplitude at frequencies $\nu \la 2.7$~GHz.  ESEs are possibly 
related to large radio continuum loops in the Milky Way 
(\citeauthor{fiedler94a} \citeyear{fiedler94a};~\citeauthor{fiedler94b} 
\citeyear{fiedler94b}).  While there are no Milky Way Galactic loops near 
GLG011, the source does lie on the edge of a very large SNR/super-bubble in 
M31.  It is possible that the observed variability is caused by this
structure, although more information is needed.  If this was the case,
than the intrinsic size of GLG011 must be very small and therefore it
could not be a PWN.

\paragraph {Pulsar Candidates}
\label{pulscans}
Of the three GLG sources identified as pulsar candidates, GLG014 and GLG036 
fall under the definition of Compact Extremely Steep--Spectrum (CESS) sources, 
which are discussed in detail in Section \ref{cess}.  While these sources are 
steep spectrum sources located inside the optical disk of M31, they
are most likely not radio pulsars in M31 because, if so, they would be
several orders of magnitude more luminous than any known galactic
pulsar.  In addition, GLG036 and GLG205 are slightly extended, which
one would not expect from a pulsar.  Additionally, GLG205 lies far off
the optical disk of M31, not expected for a radio pulsar in M31.  We
conclude that we did not detect any pulsars in M31, not surprising
given the sensitivity of our survey.

\subsubsection{HII Regions}
\label{hiiregs}
HII regions, the regions of ionized material that surround new O or B stars, 
are a major component of the radio emission of a normal galaxy.  The radio 
spectrum emission of an HII region is flat at higher radio frequencies 
($\alpha \sim -0.1$) but turns over a lower frequencies as a result of 
free--free absorption.  For roughly half of all Milky Way HII regions this 
turnover occurs at $\nu_t>325$~MHz \citep{shaver83}, so an HII region might 
appear as a flat--spectrum source or a spectral turnover source in our survey.
  To account for these two possibilities, a GLG source was classified as a HII 
region if it was inside the optical disk of M31 ($R_{M31} \leq 27~\mbox{kpc}$),
 not a Low--Frequency Variable or High--Frequency Variable (as defined in 
Section \ref{variables}), and was either:
\begin{itemize}
\item The counterpart of an optical HII region and a flat spectrum source,
\item A spectral turnover source with a flat spectrum, as defined in Section 
\ref{specclass}, that has not been identified as a SNR, or 
\item A flat spectrum source, as defined in Section \ref{specclass}, with no 
spectral turnover ($\mid$$\varphi$$\mid$$\leq\sigma_{\varphi}$) that has not 
been identified as a SNR.
\end{itemize}
Two sources, GLG016 and GLG088 satisfy the second criteria, while none satisfy 
the first or third.  Unfortunately, a flat--spectrum radio galaxy (FSRG) would 
also satisfy the second criteria.  The presence of Far-IR emission,
while not seen in all HII regions, would indicate that a source is a
HII region since FSRGs are not strong FIR emitters.  However, as
discussed in Section \ref{fir}, it is not currently possible to
reliably associate FIR objects with the radio sources detected in this
survey.  As a result, we did not detect any strong HII region
candidates in the GLG source list.  Since most Milky Way HII regions
when placed at the distance of M31 are much fainter than the detection
limit of this survey, this is not unexpected.

\subsection{Milky Way Sources}
\label{mwsrcs}
We believe that we have found two types of Milky Way radio sources in the 
GLG catalog, radio stars and Planetary Nebulae.   Radio emission has previously
 been detected from a wide variety of stars: from pre-main sequence K--stars 
\citep{kutner86} to O-B stars \citep{rauw02}.  The cause of the radio emission 
is equally varied -- the radio emission of O-B stars is expected to be 
dominated by free-free emission from stellar winds \citep{rauw02}, while radio 
emission from pre-main sequence stars is believed to be the result of 
chromospheric emission \citep{kutner86}.  In addition, radio emission has been 
detected from binary systems \citep{richards93}, and possibly from stars 
orbiting a microquasar \citep{marti98}.

A GLG source was categorized as a radio star if it had a bright optical/IR 
counterpart or had IR colors consistent with that of a star.  Three GLG sources
 met this criteria: GLG023 -- which has an $I\approx13$ counterpart in 
the MagOpt catalog while most sources in this catalog have $I=18-20$, GLG097
 -- which has the IR colors of a star, and GLG116 -- which has a bright PSC 
counterpart ($K\approx5$ as opposed to $K\sim13-15$).  R--band images of these 
sources is shown in Figure \ref{redgd}.  All three have non-thermal radio 
spectrum ($\alpha$=$-0.89$, $-0.92$, and $-0.79$,~respectively), consistent 
with the processes mentioned above.  GLG023 has an optical counterpart in the 
\citet{berk88} catalog of bright stars in the field of M31, which has a B 
magnitude of 15.04$\pm$0.19.  This is similar to the B magnitude of its 
counterpart in the MagOpt catalog ($14.9\pm0.04$), implying the source is not 
variable and possibly ruling out a binary explanation for the radio emission.
  The spectral class of this star is unknown.  GLG097 is also in a catalog of 
reference stars in the direction of M31 \citep{shokin98}, and the spectral 
class of this star is also unknown.  GLG116 is likely the radio 
counterpart of SAO 36591, a K0 star \citep{roeser88}.  Radio emission from K0 
type stars have often been associated with Algol--Type Binaries, which often 
are X--ray sources and typically have radio emission on the order of a few mJys
 at $\nu \sim 5$~GHz \citep{richards93}.  GLG116 has no X--ray counterpart, 
and has $S_{5400}=22$~mJy, brighter than that of most observed Algol-type 
binaries but not all, e.g Algol itself \citep{richards93}.

In addition to detecting radio stars, we believe that GLG347 may be a Planetary
 Nebula in the Milky Way.  GLG347 is located in the SE corner
 of our FOV and has counterparts in the 5th Cambridge Survey (5C3.152), 3rd 
Bologna Survey (B3 0043+398), and the Miyun 232 MHz survey 
(MY 004309.7+394955.8).  On the basis of these detections, it has previously 
been identified as a radio galaxy \citep{simbad}.  However, as shown in Figure 
\ref{glg347pic}, located to the SE of this source is a region of increased 
emission not classified as a source by $\mathcal{SFIND}$.  Taken together, 
these regions suggest a shell of diameter $\sim$45\arcsec~and a flux density 
$S_{325}$=18.5~mJy.  This is smaller and fainter than most galactic SNRs 
\citep{greensnrcat}, and at $b \approx -23^{\circ}$, much further off 
the galactic plane that most Milky Way SNRs, so we do not believe that is a 
SNR.  However, this source has a diameter, morphology and flux density similar 
to that of planetary nebulae, and this is the basis of its classification 
\citep{pneb}.  There is an IR source in the center of the radio emission, 
shown in Figure \ref{glg347pic}, with the IR colors of a star which could be 
the source of the Planetary Nebula.

\subsection{Background Sources}
\label{extragal}
In this section, a ``background'' source is one neither in M31 or in the Milky 
Way.  The population of background sources is dominated by galaxies whose 
radio emission is powered by a central AGN (primarily FRI and FRII radio
 galaxies) -- particularly sources with $S_{325} > 10$~mJy,  and galaxies 
whose radio emission is dominated by synchrotron emission from relativistic 
electrons and free--free emission from HII regions (normal galaxies, 
\citeauthor{condon} \citeyear{condon}) -- expected to dominate at 
$S_{325} < 10~\mbox{mJy}$ (Carole Jackson, private communication 2002).  
Identifying these sources is crucial in determining whether or not a source is 
in M31 or the Milky Way, since, as shown in Section 4 of Paper 1, 
``background'' sources dominate the GLG source list.

Even though this class of sources is not the primary motivation nor the primary
 focus of this paper, many of these are interesting objects: we detected a BL 
Lac, a galaxy merger, several High--Frequency and Low--Frequency Variable 
sources, and have identified a Giant Radio Galaxy (GRG) candidate and several 
High-z Radio Galaxy  (HzRG) candidates.  This section will first briefly 
describe the sources which we believe are typical background sources (FRI, 
FRII, and normal galaxies) and then discuss the more exotic background sources 
in detail.

\subsubsection{FRI \& FRII Radio Galaxies and Radio Jets}
\label{agn}
The most common source with $S_{325} \stackrel{>}{\sim} 5$~mJy are galaxies 
whose radio emission is dominated by a central super-massive black hole 
(Active Galactic Nuclei; AGN).  These sources typically fall into three broad 
morphological categories:
\begin{description}
\item{\bf{FRI:}}  A galaxy whose radio emission appears core dominated 
\citep{fr}.  A GLG source was categorized as an FRI if it was either:
\begin{itemize} 
\item resolved and the radio emission is dominated by a single, central 
component (Criterion A in Table \ref{frIcan}).  A resolved PWN may have 
such a morphology, but we do not expect to be able to resolve a PWNe in M31.
\item an unresolved or elongated source with a counterpart in the ROSAT 2nd 
PSPC survey categorized as a ``Galaxy" source which does not have a good IR 
counterpart (Criterion B in Table \ref{frIcan}).  As shown in Section 
\ref{ir}, the detected IR emission from GLG sources is too bright for the 
source to be an FRI/FRII explanation.
\item an unresolved or elongated source outside the optical disk of M31 
with an X--ray counterpart and no IR counterparts (Criterion C in Table 
\ref{frIcan}).
\end{itemize}
As mentioned in Section \ref{ir}, no GLG source with a counterpart in the 
2MASS catalogs had a K--magnitude consistent with that of FRI.  Table 
\ref{frIcan} lists the nine GLG sources determined to be FRIs.  One of these 
is also a High--Frequency Variable, as defined in Section \ref{variables}, and 
two are also High-z Radio Galaxy candidates, as defined in Section \ref{hzrgs}.
  The IR counterpart of GLG122 is a star, and most likely unrelated to the 
radio source.  All but two of these sources have X--ray counterparts.  This 
Table is by no means complete since most likely a vast majority of unresolved 
sources detected outside of the optical disk of M31 are FRIs, it only lists 
those sources for which there is additional evidence that they are FRIs.

\item{\bf{FRII:}}  A galaxy whose radio emission is dominated by radio jets 
emitted from the central black hole, and often more luminous than FRIs 
\citep{fr}.  FRIIs may or may not have a ``core'' component that corresponds 
to the central AGN.  A GLG source was categorized as an FRII if it was 
resolved and had the morphology described above.  Table \ref{frIIcan} lists 
the twenty-three GLG sources determined to be FRIIs.  GLG015 and GLG078 are 
also classified as anomalous sources because of H$\alpha$/IR emission detected 
near these sources.  GLG078, GLG129, GLG187, and GLG296 are also HzRG 
candidates (Section \ref{hzrgs}), and GLG187, GLG296, and GLG358 are 
Low--Frequency Variables (Section \ref{variables}).  GLG220 and GLG212 form a 
radio triple, shown in Figure \ref{radtrip}.  Seven FRIIs have 2MASS 
counterparts, and of these four have the IR colors of a normal galaxy.  For 
these four sources we believe that the radio emission is from the central AGN 
but the IR emission is from the rest of the host galaxy.  The IR counterparts 
of the other three are most likely coincidental.

\item{{\bf Radio Jets:}}  Radio jets are most commonly emitted from AGN, but 
tend to be larger in angular size than FRIIs which is why they are in a 
separate category.  A GLG source was classified as having radio jets purely on 
a morphological basis.  Figure \ref{jetspic} shows the five radio jets 
discovered in this survey: GLG031/GLG033, GLG045/GLG051, GLG054/GLG059, and 
GLG266/GLG269, and GLG270/GLG271/GLG275.  Between GLG031 and GLG033 is an 
elongated 20 cm source in the Braun Catalog (Braun 3), not seen in our 325 MHz 
image, oriented perpendicular to GLG031/GLG033, that possibly is the host 
galaxy.  The bright point sources 
in GLG031 and GLG033 are most likely hot spots in the jet \citep{carilli98}. 
GLG045/GLG051 is a one--sided jet, where GLG045 is the central core while 
GLG051 is the jet itself.  The bright point source 
in GLG051 is most likely the hot spot produced at the termination point of the 
jet \citep{carilli98}.  In the case of GLG054/GLG059, GLG054 has the morphology
 of a radio jet, and its extended emission is oriented towards GLG059. The 
GLG266/GLG269 system also does not have the classical morphology of a radio 
jet -- GLG269 is a steep spectrum ($\alpha \sim -2$) complex source with the 
morphology of a one-sided jet, while GLG266 is a HzRG candidate (Section 
\ref{hzrgs}) along the jet axis of GLG269. The GLG270/GLG271/GLG275 is a 
classic radio jet, with GLG271 most likely the central engine that produces 
GLG270 and GLG275.
\end{description}
 
\subsubsection{Normal Galaxies}
\label{galaxies}
While Section \ref{agn} identified background AGN, this section identifies 
background normal galaxies -- those whose radio emission is not dominated by 
a central AGN but from synchrotron emission emitted by relativistic electrons 
(``cosmic rays'') in SNRs and in diffuse gas interacting with the galaxy's 
magnetic field and free--free 
emission produced by HII regions.  Normal radio galaxies have optical 
morphologies across the entire Hubble diagram \citep{condon}.  A GLG source 
was determined to be a normal galaxy if it satisfied at least one of the 
following criteria:
\begin{description}
\item[Criterion I:] Has an 2MASS PSC counterpart with the IR colors of
  a normal galaxy (as discussed in Section \ref{ir}).
\item[Criterion II:] Has a counterpart in the 2MASS ESC catalog.
\item[Criterion III:] Has a counterpart in the {\it ROSAT} 2nd PSPC
  survey that was classified as a galaxy and in either the 2MASS ESC
  or 2MASS PSC.
\end{description}
Table \ref{rgcan} lists the GLG sources identified as radio
counterparts to normal galaxies.  As shown in this table, several of
these sources fall into other categories.  In these cases, we believe that the
observed radio and IR emission come from different parts of the source
-- the radio emission from the central AGN and the IR emission from
the rest of the galaxy.  GLG253 is located in the center of a z=0.293
galaxy cluster identified by the {\it XMM--Newton} satellite \citep{galclust}.

Two of these sources -- GLG050 and GLG203 -- are ``spectral curvature'' sources
 as defined in Section \ref{specclass}.  For GLG203 there is no evidence for 
ionized material along the line of sight, so the observed spectral turnover is 
most likely intrinsic.  Intrinsic low--frequency spectral turnover has been 
seen in many normal galaxies, and can be explained by a clumpy, non--thermal 
plasma with electron temperatures $200~{\rm K} \leq T_e \leq1500~{\rm K}$ and 
electron density $n_e=0.07-0.10$~cm$^{-3}$ \citep{im90}.  For a typical galaxy,
 the optical depth at 325~MHz of such a medium ($n_e=0.1$~cm$^{-3}$ and 
$T_e=100$~K) is $\sim$0.002, an order of magnitude lower than that observed 
in GLG203.  However, the EM of these galaxies is similar to that of the 
starburst galaxy M82 (\citeauthor{allen} \citeyear{allen};~\citeauthor{wills} 
\citeyear{wills}).  The cause of the negative spectral curvature observed in 
GLG050 is unclear.

\subsubsection{High-z Radio Galaxies}
\label{hzrgs}
Three separate criteria, all sensitive to different kinds of High-z Radio 
Galaxies (HzRGs), were used to identify HzRG candidates.  The first criterion 
used the fact that Ultra--Steep Spectrum (USS) sources, defined here to be 
$\alpha \leq -1.3$, have been determined to be efficient tracers of HzRGs 
(\citeauthor{hzrg} \citeyear{hzrg}; \citeauthor{pedani03} \citeyear{pedani03}).
  This is believed to be an intrinsic property of HzRGs since the generic 
spectrum of a radio galaxy lobe, assumed to behave like an optically thin 
synchrotron source, steepens in spectral index by $\sim$0.5 around the 
``bend'' frequency $\nu_{b}$ as one goes from low to high frequencies due to 
synchrotron losses.  The observed $\nu_{b}$ is proportional to $({1+z})^{-1}$, 
so at higher redshifts the bend--frequency occurs at lower frequencies, causing
 the object to have a steep spectral index when observed at frequencies above 
the redshifted bend frequency.  Inverse--Compton scattering off the CMB also 
contributes to the steeping of the spectrum of HzRGs, since losses due to this 
process increase as $(1+z)^4$ \citep{krolik91}.  Using this information, a 
source was identified as a HzRG if it was resolved with the morphology of an 
FRI or an FRII and has $\alpha \leq -1.3$.  The four sources that met this 
``USS'' criterion are listed in Table \ref{hzrgcan}, implying a source density 
of $\sim$1700~sr$^{-1}$.  This is more than 10$\times$ higher than that 
measured in a comparison between the WENSS and NVSS surveys \citep{hzrg}.

The second criterion is designed to detect two subclasses of potential HzRGs 
which have a more compact morphology -- Gigahertz Peaked-Spectrum (GPS) 
sources and Compact Steep Spectrum (CSS) sources \citep{kaplan00}.  Both 
classes are extremely powerful, compact\footnote{GPS sources are $\la$1~kpc in 
size while CSS generally are $\sim$1--20~kpc in size \citep{odea98}.} sources 
with radio spectra that turnover at low--frequencies.  For GPS sources, 
this turnover occurs at $\nu_t \sim$1~GHz, while for CSS sources this turnover 
generally occurs at $\nu_t \la$500~MHz \citep{odea98}.  At frequencies above 
$\nu_t$, the spectral index of GPS sources is typically between 
$\alpha \approx -0.5$ and $\alpha \approx -0.9$  \citep{odea98}, while for 
CSS sources it is between $\alpha \approx -0.5$ and $\alpha \approx -1.5$ 
\citep{fanti90}.  The spectral turnover is most likely due to synchrotron 
self--absorption, but free--free absorption may also play a role 
\citep{odea98}.  A GLG source was determined to be a CSS or a GPS source 
if it is a ``spectral curvature'' source with $\varphi >$~0 and spectral index
 $\alpha <$-0.5 that has not been classified as a ``normal'' galaxy (see 
Section \ref{galaxies} for details).  Since only measurements at three 
frequencies are available, we are unable to distinguish between CSS and GPS 
sources, though a visual inspection of the spectra of these sources implies
that most of them are CSS sources.  The sources that met this
``CSS/GPS'' criterion are listed in Table \ref{hzrgcan}.

High--resolution ($\theta_{res} < 1$\arcsec) VLA images of CSS sources have 
shown that they are often doubles, triples, or even have jets associated with 
them (\citeauthor{fanti90} \citeyear{fanti90}; \citeauthor{athreya97} 
\citeyear{athreya97}).  CSS sources with extended emission almost always have 
cores with a flat or inverted spectrum ($-0.5<\alpha<1.0$) and extended 
emission with a spectral index of $-1.0 < \alpha < -0.7$ \citep{athreya97}.  
If these two components are of comparable strength than it is possible that at 
higher frequencies ($\nu \sim$1~GHz) the ``flat--spectrum'' core dominates 
while at lower frequencies ($\nu \sim$300~MHz) the steeper--spectrum extended 
emission dominates, resulting in a spectrum with $\varphi <$0.  CSS sources of 
this nature were selected from the GLG sample by requiring that the source be 
a ``spectral curvature'' source outside the optical disk of M31 with 
$\varphi <$~0.  The sources that met this ``CSS'' criterion are also
listed in Table \ref{hzrgcan}.

Table \ref{hzrgcan} lists the GLG sources that met at least one of these
criteria.  GLG296 was previously identified as a HzRG candidate based on its 
properties in the WENSS and NVSS catalogs \citep{hzrg}, and several 
of these objects fall under other classifications as well.  Follow up
radio and optical observations are needed to determine if these
objects are truly High-z Radio Galaxies.

\subsubsection{Galaxy Merger}
\label{glg247}
Galaxy mergers are powerful radio sources because the collision triggers star 
formation, which in turn creates radio--bright sources such as HII regions and 
SNRs.  Radio observations of galaxy mergers, such as the ``Antennae'' galaxies 
(NGC 4038/4039), have discovered that the radio emission from these sources is 
concentrated in the region of brightest optical emission \citep{neff00}.  
GLG247 is classified as a galaxy merger as a result of its radio morphology, 
as well as its relationship with its IR, optical, and X--ray counterparts, 
shown in Figure \ref{glg247pic}.  GLG247 has a relatively steep spectrum, 
$\alpha=-0.87$, so it is possible that the radio population is dominated by 
SNRs like that of the ``Antennae'' galaxies \citep{neff00}, but higher 
resolution radio images are needed to determine the source of the radio 
emission.  GLG247 could also be an ``X--shaped'' radio galaxy \citep{leahy84}.

\subsubsection{BL Lacs}
\label{bllac}
BL Lacs -- believed to be the radio emission from a jet aligned with our 
line-of sight emitted by a super-massive black hole -- are some of the most 
powerful radio sources known and have been detected throughout the EM spectrum,
 from the radio up to TeV energies \citep{tagliaferri03}.  The X--ray 
counterpart of GLG105
 was previously identified as a BL Lac candidate in the Einstein Slew Survey 
\citep{perlman96}.  In our survey, GLG105 is a a complex source with a 
cometary morphology (shown in Figure \ref{glg105pic}) and a radio spectrum 
with 
spectral index $\alpha=-0.74\pm0.10$ and no curvature, but shows 
variability at high--frequencies (see Section \ref{variables} for more 
details).  Similar variability has been observed in other BL~Lacs (e.g. 
\citeauthor{padrielli87} \citeyear{padrielli87}).  GLG105 has optical, IR, and 
X--ray counterparts, 
and its broadband SED (radio to optical) is shown in Figure \ref{glg105pic}.  
This spectrum does not fit general spectral models of BL~Lacs, in which a 
power--law extrapolation of the radio emission over-predicts 
the observed optical/IR emission \citep{bllacsed}.  This discrepancy can be 
explained by variability between the different epochs of the observation, but 
is interesting 
nevertheless.  Additionally, H$\alpha$ emission has been detected near the 
``tail'' of GLG105 (shown in Figure \ref{glg105pic}), uncharacteristic
 of BL~Lac objects.  However, this source has a $log(f_x/f_V)=1.6$, consistent
 with that of other known BL~Lacs \citep{maccacaro88}.  Since GLG105 is located
 within the optical disk of M31, it is possible that it is a very luminous 
($L_X=2.4 \times 10^{38}$~ergs s$^{-1}$; \citeauthor{rosat} \citeyear{rosat}) 
X--ray source in M31, but is extremely unlikely due to the rarity of such 
objects.

\subsubsection{Giant Radio Galaxies}
\label{grg}
As mentioned in Section \ref{agn}, some central AGN are known to produce 
relativistic jets which -- if powerful enough -- inflate a ``cocoon'' that 
expands first into the ISM of the host galaxy, and then into the surrounding 
Intergalactic Galactic Medium (IGM).  The magnetic fields in these cocoons 
produce synchrotron emission, which is observed in the form of radio lobes.  A 
Giant Radio Galaxy (GRG) is a radio galaxy whose lobes span a projected 
distance of $>1$~Mpc.  Since radio lobes expand with time, GRGs must be very 
old ($>10^8$ years), and located in under-dense regions relative to other, 
more compact, FRIs and FRIIs.  GRGs are, as a result of their large size, very 
powerful probes of the IGM \citep{schoenmakers01}.

We believe that GLG242 and GLG260, shown in Figure \ref{grgpic}, located 
4\arcmin~away from each other, together comprise a Giant Radio Galaxy 
(GRG).  (According to the search criteria of \citet{schoenmakers01}, the
 radio lobes need to span 5\arcmin~in order for the source to be considered a 
GRG candidate.  However, since the redshift of this object is unknown, the 
physical distance between these two objects could be $>1$~Mpc.)  GLG242 is a 
complex source which composed of a point source with an attached region of 
extended emission oriented in the general direction of GLG260, an unresolved 
source in our nomenclature but which, under close examination, has extended 
emission in the direction of GLG242.  GLG242 has a flux density of 
$S_{325}=254$~mJy and a spectral index $\alpha=-1.07\pm0.42$, while GLG260 is 
somewhat weaker ($S_{325}=176$~mJy) and has a flatter spectrum 
($\alpha=-0.81\pm0.031$).  These differences are similar to that seen in other 
GRGs (Table 3 of \citeauthor{schoenmakers00} \citeyear{schoenmakers00}).  When 
combined, these sources have a total flux density of $S_{325}\sim$430~mJy -- 
well within the flux range of known GRGs (Figure 1a of 
\citeauthor{schoenmakers01} \citeyear{schoenmakers01}), and a composite 
spectral index $\alpha_{325}^{1400}=-0.94$ -- steeper than all but a handful 
of known GRGs (Table A4 of \citeauthor{schoenmakers01} 
\citeyear{schoenmakers01}).

A more detailed analysis of these sources is hampered by the lack of 
information about their host galaxy.  As seen in the 325~MHz and the 1.4~GHz 
(NVSS) radio image of the GLG242/GLG260 (shown in Figure \ref{grgpic}) system, 
there is no radio source between them.  While most host galaxies of GRGs have 
$S_{325}>3$~mJy (the 5$\sigma$ RMS limit in the area between these sources), 
several host galaxies of GRGs have 325~MHz flux densities below this limit so 
it is not inconceivable that this is the case for GLG242/GLG260 (Table
 A3 of \citeauthor{schoenmakers01} \citeyear{schoenmakers01}).  In addition, 
there is no obvious optical/IR candidate for the host galaxy.  There is an IR 
point source located $\sim$20\arcsec~south of the center of GLG242, but 
this source has the IR colors of a star (see Section \ref{ir} for details), so 
it is probably unrelated.  GLG260 has no counterparts in the Far-IR, IR, 
optical, or X--ray catalogs/images searched.  A J--band image of GLG242/GLG260 
(also shown in Figure \ref{grgpic}) reveals the existence of a faint 2MASS 
source (2MASS:00465464+4033383) on the line that connects GLG242 and GLG260, 
but based on its IR-colors and its optical properties in the USNO B-1.0 
catalog \citep{usno}, it is most likely a star.

\subsubsection{Variable Sources}
\label{variables}
Flux variability -- whether at low ($\nu<1$~GHz) or high ($\nu\geq1$~GHz) 
frequencies, or on short ($t_{var}<$1~yr) or long ($t_{var}\geq$1~yr) 
timescales -- provides a wealth of information about the source itself as well 
as on the intervening ISM.  Studies of variable radio sources have that they 
typically vary on timescales longer than a year \citep{gregorini86}, and that
 source which vary at low frequencies ($\nu \la 1~$GHz) have different 
 properties (spectral index, galactic latitude, variability magnitude) than 
those which vary at high frequencies ($\nu \ga 1~$GHz) 
(\citeauthor{cawthorne85} \citeyear{cawthorne85}; \citeauthor{gregorini86} 
\citeyear{gregorini86}; \citeauthor{rys1990} \citeyear{rys1990}).

It is currently believed that variability at low and high frequencies are 
caused by two separate processes:
\begin{itemize}
\item A process intrinsic to the source -- which is almost always an active 
galactic nucleus (AGN) of some sort -- that dominates at higher frequencies 
but can cause variability at lower frequencies.
\item A process extrinsic to the source, caused by material in the Milky Way, 
that dominates variability at low frequencies -- Interstellar Scintillation 
(ISS).
\end{itemize}
This intrinsic process is believed to occur in a majority, if not all, flat 
spectrum sources -- most likely QSOs or BL Lacs \citep{rys1990} -- with 
$S_{1400} \geq 10$~mJy \citep{carilli2003}.  The extrinsic process is believed 
to be Refractive Interstellar Scintillation (RISS), the result of radio waves 
from the extragalactic source passing through a turbulent region of the ISM 
\citep{rickett84}, is also used to explain slow flux variations observed in 
Galactic pulsars \citep{cordes98}.

Using data available from the WENSS survey, we searched for low-frequency 
variables (LFVs) in the GLG source list.  Since the WENSS source list was 
derived from observations conducted in 1991--1993 \citep{wenss} while the GLG 
sources were observed in 2000, we are sensitive to LFVs that vary on a 
$\sim$7--10 year timescale. In addition, since the WENSS FOV includes the 
entire GLG FOV, we can detect LFVs across the entire FOV and see if the ISM of 
M31 plays a role.  A GLG source was identified as LFV if it satisfied one of 
the following two criteria:
\begin{description}
\item [Criterion 1:] The GLG source had a good match in the WENSS catalog, and 
had:
\begin{equation}
\Delta S_{325} \geq 5 \times \sqrt{(\sigma_S^{GLG})^2+(\sigma_S^{WENSS})^2}
\end{equation}
where $\Delta S_{325}$ is the absolute value of the difference in 325 MHz 
flux densities between the two sources, $S_{325}^{GLG}$ is the measured flux 
density of the source in the GLG catalog, $S_{325}^{WENSS}$ is the measured 
flux density of the corresponding source in the WENSS catalog, 
$\sigma_S^{GLG}$ is the error of $S_{325}^{GLG}$, and $\sigma_S^{WENSS}$  is 
the error of $S_{325}^{WENSS}$
\item[Criterion 2:]  The GLG source had no counterpart in the WENSS catalog, 
and:
\begin{equation}
S_{325}^{GLG} \geq 30~\mbox{mJy} + 5\times \sigma_S^{GLG}
\end{equation}
where $S_{325}^{GLG}$ and $\sigma_S^{GLG}$ are defined as above.  This criteria
 looks for GLG sources that should have been, but were not, detected in the 
WENSS survey, which has a completeness limit of 30~mJy \citep{wenss}.
\end{description}
Table \ref{lfvcan} lists the 10 GLG sources that were classified as LFV, and 
for comparison purposes, a non-LFV GLG source (last line of Table).  The 
columns are as follows:
\begin{description}
\item[Column 1:] GLG name
\item[Column 2:] Morphology Type
\item[Column 3:] $R_{M31}$, as defined in Section 4.1 in Paper I, in kpc.
\item[Column 4:] Radio Spectral Index $\alpha$, calculated using Equation 7 in
 Paper I.
\item[Column 5:] Spectral curvature parameter $\varphi$, calculated using 
Equation 7 in Paper I.
\item[Column 6:] Name of WENSS counterpart.
\item[Column 7:] Distance between GLG source and WENSS counterpart in 
arcseconds.
\item[Column 8:] $S_{325}^{GLG}$, as defined above, in mJy.
\item[Column 9:] $S_{325}^{WENSS}$, as defined above, in mJy
\item[Column 10:] Variation amplitude 
$\frac{\Delta S_{325}}{\overline{S_{325}}}$, where $\Delta S_{325}$ is defined
 above and $\overline{S_{325}}={S^{GLG}_{325}+S^{WENSS}_{325}}$/2
\end{description}

Table \ref{lfvcanhf} lists the 1.4 GHz properties of these LFVs.  The columns 
are as follows:
\begin{description}
\item[Column 1:] GLG name
\item[Column 2:] Name of 1.4 GHz counterparts
\item[Column 3:] Distance from GLG source to 1.4 GHz counterparts, 
in arcseconds.
\item[Column 4:] Flux density of 1.4 GHz counterparts in mJy.
\item[Column 5:] $\overline{S_{1400}}$, the average flux density of the 
1.4~GHz counterparts, in mJy.
\item[Column 6:] $\frac{\Delta S_{1400}}{\overline{S_{1400}}}$, where 
$\Delta S_{1400}$ is the difference in flux density of the 1.4 GHz 
counterparts, and $\overline{S_{1400}}$ is defined below.
\end{description}

A total of 95 GLG sources\footnote{Extended sources with no radio 
counterparts were excluded from this sample since they are, most likely, false 
detections.} either had a good counterpart 
in the WENSS catalog, or had $S_{325}^{GLG} \geq$30~mJy with no counterparts 
in the WENSS catalog.  Of these, $\sim$11\% of them are LFVs, a significantly 
higher percentage than that observed in previous LFVs searches.  Additionally, 
the observed variability is higher than in previous studies 
\citep{stinebring00}, especially given the high Galactic latitude of M31 
($|b|=22^{\circ}$).  In addition, a large number of the LFVs in the GLG sample 
are ``steep'' spectrum ($\alpha \leq -0.5$) and/or resolved sources, different 
from previous LFV studies \citep{riley93}.  For five of the steep--spectrum 
LFVs it was possible to calculate $\varphi$, and two of them (GLG200 and 
GLG340) show no spectral curvature -- unlike previously detected steep 
spectrum LFVs \citep{riley93}.  However, it is possible that lower frequency 
observations of these two sources would show spectral flattening.  Of the three
 resolved LFVs -- two of which are FRIIs (GLG187 and GLG296) and one 
is an Extended source (GLG009) -- only GLG009 and GLG296 have sufficient 
information at 1400~MHz to determine if they are high-frequency variable (HFVs)
 or have spectral curvature.  GLG009 is a HFV, surprising because most HFVs 
are compact \citep{rys1990} and there doesn't appear to be a compact source 
inside GLG009.  GLG296 is not a HFV, but does have spectral turnover similar 
to that previously detected in LFVs.  While the larger resolution of 
the WENSS survey can explain a higher flux density in this survey because the 
recorded flux is inflated by extended emission in the region resolved out in 
the GLG survey, over half of the detected LFVs have higher fluxes in the GLG 
survey than the WENSS survey.  However, the fluxes of GLG009  and GLG296 is 
lower in the GLG survey than in the WENSS survey so these source might not be 
variables.

By using the fluxes of a GLG source's 1.4 GHz counterparts, we were able to 
determine if a source is a HFV.  The criteria were:
\begin{description}
\item[Criterion 1:] The GLG sources has two good 1400~MHz counterparts, and 
the $S_{1400}$ of the two counterparts differed by more than 
5$\sigma_{\rm diff}$, where $\sigma_{\rm diff}$ is the error in the difference.
\item[Criterion 2:] The GLG sources has more than two good 1400~MHz 
counterparts, and for at least one of these counterparts:
\begin{equation}
|S_{1400}-\overline{S_{1400}}| \geq 5\times \sqrt{(\sigma_S)^2 + 
(\sigma_{\bar{S}})^2}
\end{equation}
where $\sigma_S$ is the error in $S_{1400}$ of the 1400~MHz counterpart in 
question and $\sigma_{\overline{S}}$ is the error of $\overline{S_{1400}}$, the
 weighted average of all the 1400~MHz flux densities.
\end{description}
Table \ref{hfvcan} for comparison lists the six HFVs identified in the GLG 
source list, and a 
non-HFV GLG source (last line of Table).  The columns are as follows:
\begin{description}
\item[Column 1:] Name of GLG source
\item[Column 2:] Radio spectral index, $\alpha$, as calculated in Section 
4.3 in Paper I.
\item[Column 3:] Spectral curvature parameter, $\varphi$, as calculated in 
Section 4.3 in Paper I.
\item[Column 4:] Names of 1.4 GHz counterparts.
\item[Column 5:] Distance from 1.4 GHz counterparts to GLG source, in 
arcseconds.
\item[Column 6:] Flux density of 1.4 GHz counterparts, $S_{1400}$, in mJy.
\item[Column 7:] Variation amplitude 
$\frac{\Delta S_{1400}}{\overline{S_{1400}}}$, where $\Delta S_{1400}$ is 
the difference between the minimum and maximum $S_{1400}$ measurements and 
$\overline{S_{1400}}$ is defined above.
\end{description}

Table \ref{hfvcanlf} lists the 325~MHz properties of these HFVs.  The 
columns are as follows:
\begin{description}
\item[Column 1:] Name of GLG source.
\item[Column 2:] Morphology type.
\item[Column 3:] $R_{M31}$, defined in Section 4.1 in Paper I, in kpc.
\item[Column 4:] Name of WENSS counterpart.
\item[Column 5:] Distance between GLG source and WENSS counterpart in 
arcseconds.
\item[Column 6:] Flux density of source in GLG observation, $S_{325}^{GLG}$ in 
mJy.
\item[Column 7:] Flux density of WENSS counterpart, $S_{325}^{WENSS}$ in mJy.
\item[Column 8:] Average 325~MHz flux density of source, $\overline{S_{325}}$,
 in mJy.
\item[Column 9:] 325~MHz Variation amplitude, 
$\frac{\Delta S_{325}}{\overline{S_{325}}}$, where $\Delta S_{325}$ and 
$\overline{S_{325}}$ are defined above.
\end{description}

There are 120 sources in the GLG source list with two or more good 1400~MHz 
counterparts, and only six (5\%) of them fit the above criteria for a HFV.  
This percentage is slightly lower than previous searches (e.g. 
\citeauthor{rys1990} \citeyear{rys1990};~\citeauthor{carilli2003} 
\citeyear{carilli2003}), but the amount of variation observed is similar 
\citep{rys1990}.  As mentioned earlier, only GLG009 is both a LFV and HFV.  
Two of the six HFVs were resolved -- GLG009, and the BL Lac candidate GLG105
 which was described in Section \ref{bllac}.  Of the four unresolved
 HFVs, only one of them is a flat--spectrum source (GLG011).  Only one of the 
steep--spectrum HFVs (GLG004) showed any appreciable spectral turnover, and 
only at the 2$\sigma$ level.

To determine if M31 has an effect on low--frequency and/or high--frequency 
variability, we plotted $\frac{\Delta S}{\overline{S}}$ at both 325 and 1400 
MHz as a function of $R_{M31}$, seen in Figure \ref{varpics}.  There is no 
observed correlation between M31 and flux variation at either 325 or 1400 MHz. 

To summarize, in the GLG source list we detected 10 LFVs and seven HFVs.  A 
higher percentage of LFVs were found in the GLG survey than in other LFV 
searches, most likely the result of this survey's sensitivity to fainter 
sources.  However, the variability is higher than seen in previous LFV surveys,
 and much higher than expected for the Galactic latitude of M31.  Many of these
 have steep spectra, unlike in other surveys, and several were resolved.  
However, the HFVs detected in the GLG survey have similar properties seen in 
other HFV searches.  Lastly, flux density variations at both 325 and 1400 MHz 
appear to be uncorrelated with structures in M31.

\subsection{``Anomalous'' Sources}
\label{anomalous}
In this section, we discuss sources with spectral, morphological, and/or 
multi-wavelength properties we do not understand.  

\subsubsection{Compact, Extremely Steep Spectrum Sources}
\label{cess}
Table \ref{cesscan} lists the 22 GLG sources which we consider to be Compact, 
Extremely Steep Spectrum (CESS) sources -- sources that are either ``U'' or 
``El'' in our radio images, not classified as HFVs, LFVs (see Section 
\ref{variables} for details), or as HzRGs (see Section \ref{hzrgs} for 
details), and have a spectral index $\alpha \leq -1.6$.  Images of these 
sources are shown in Figure \ref{cesspics}.  Only four of these sources have 
been detected at other wavelengths -- GLG014 has an X--ray counterpart, GLG036 
has an optical counterpart, and both GLG140 and GLG149 have IR counterparts, 
though these may be coincidental.  This survey is not the first time that such 
objects have been detected -- a comparison of the NVSS survey and the 365~MHz 
Texas survey catalogs found 74 sources with $\alpha \leq$-1.5 and four sources 
with spectral index $\alpha <$-2.5 \citep{kaplan00}.  This search covered 10 
sr of sky, a much larger region than our observation, but only included sources
 with $S_{365} \geq$200~mJy \citep{kaplan00}, a much higher cutoff than this 
survey ($S_{325} \sim$3~mJy) -- so the higher density of CESS sources detected 
in this survey could be a result of our sensitivity to fainter sources. 

Two possible explanations for the observed steep--spectra observed are:
\begin{description}
\item[Option \#1:] These sources have flatter spectrum than those calculated 
here, and their apparent spectrum is steep because of some intrinsic or 
extrinsic effect.
\item[Option \#2:] These sources have intrinsically steep spectra. 
\end{description} 
These options are not mutually exclusive, and neither one necessarily explains 
all the CESS sources.  

Option \#1 essentially implies that the source is variable -- either 
intrinsically
 or extrinsically.  As discussed in Section \ref{variables}, such variability 
can either occur at low frequencies (in this case $\nu=325$~MHz) or at high 
frequencies (in this case $\nu=1.4$~GHz).  As mentioned earlier, 
high--frequency variation, believed to be intrinsic to the source, occurs 
predominantly in ``flat--spectrum'' sources ($\alpha \geq -0.5$; 
\citeauthor{rys1990} \citeyear{rys1990}).  To test this explanation, we 
calculated the high--frequency 
variation needed to explain the observed steep spectrum of CESS sources 
assuming the source has an intrinsic spectral index of $\alpha=-0.5$.  Table 
\ref{cesshfv} shows the results of this analysis.  The columns of this table 
are:
\begin{description}
\item[Column 1:] Name of GLG source.
\item[Column 2:] 325~MHz flux density of GLG source in mJy.
\item[Column 3:] Extrapolated 1.4~GHz flux density of GLG source, in mJy, 
assuming a spectral index of $\alpha=-0.5$, $S_{1400}^{exp}$.
\item[Column 4:] Measured (or upper limit) 1.4~GHz flux density, in mJy, 
$S_{1400}^{meas}$.  For sources with counterparts in multiple 1400~MHz 
catalogs, this is the minimum measured value.
\item[Column 5:] Survey that this value for $S_{1400}$ comes from.
\item[Column 6:] $\frac{\Delta S_{1400}}{\overline{S_{1400}}}$ of this source,
 defined in Section \ref{variables}.
\item[Column 7:] Average percent variation ($\overline{Var_{1400}}$) of this 
source per year, equal to:
\begin{equation}
\overline{Var_{1400}}=100 \times \frac{\Delta S_{1400}}{t_{years}} 
\end{equation}
where ${\Delta S_{1400}}$ is simply $|S_{1400}^{exp} - S_{1400}^{meas}|$ and 
$t_{years}$ is the elapsed time between the GLG observation and 1.4~GHz 
observation in Column~5.
\end{description}
The required values for $\frac{\Delta S_{1400}}{\overline{S_{1400}}}$ for all 
but GLG046 and GLG152 are much higher than those observed 
in the previous HFV surveys discussed in Section \ref{variables}.  This does 
not rule out intrinsic variability as the cause of CESS sources since the 
1.4~GHz surveys used in the above calculation are much deeper than those used 
in previous HFV surveys.  However, at $S_{1400} \la 1-10$~mJy, normal 
galaxies are an increasingly dominant fraction of the radio galaxy population 
\citep{hopkins98}, and since these sources are not expected to be extremely 
variable, it is not obvious that fainter 1.4~GHz sources will have higher flux 
variability.  

Variability at low--frequencies is also a possible explanation of the CESS 
population.  
As stated in Section \ref{variables}, low--frequency variation is caused by 
two separate types of processes: intrinsic variation and interstellar 
scintillation.  If low--frequency variation is intrinsic, then the source 
would be a Group 1 LFV (See Section \ref{variables}), and previous studies of 
such LFVs have found that these sources have ``flat'' spectrum 
($\alpha \geq -0.5$) \citep{gregorini86}. Therefore, we could analyze this 
possibility in a similar fashion as we did for the high--frequency variation 
possibility -- using the available 1.4 GHz data, calculate the 325~MHz of the 
object at the time of 1.4~GHz observation, and then determine the variability 
needed to explain the 325~MHz flux density measured in the GLG survey.  Table 
\ref{cesslfv} presents the results of this analysis.  The columns of this 
table are:
\begin{description}
\item[Column 1:] Name of GLG source.
\item[Column 2:] 325~MHz flux density of GLG source in mJy.
\item[Column 3:] Measured (or upper limit) 1.4~GHz flux density, in mJy, 
$S_{1400}^{meas}$.
\item[Column 4:] Survey that this value for $S_{1400}$ comes from.
\item[Column 5:] 325~MHz flux density of GLG source, in mJy, assuming a 
spectral index of $\alpha=-0.5$, $S_{325}^{exp}$.
\item[Column 6:] $\frac{\Delta S_{325}}{\overline{S_{325}}}$ of this source,
 defined in Section \ref{variables}.
\item[Column 7:] Average percent variation ($\overline{Var_{325}}$) of this 
source per year, equal to:
\begin{equation}
\overline{Var_{325}}=100 \times \frac{\Delta S_{325}}{t_{years}} 
\end{equation}
where ${\Delta S_{325}}$ is simply $|S_{325}^{exp} - S_{325}^{meas}|$ and 
$t_{years}$ is the elapsed time between the GLG observation and 1.4~GHz 
observation in Column~5.
\end{description}
As was the case for High--Frequency variation, the variability required to 
explain the observed steep--spectra of the CESS sources if they are flat 
spectrum, intrinsically varying LFVs is much higher than that seen in previous 
LFV studies (e.g. \citeauthor{rys1990} \citeyear{rys1990}; \citeauthor{riley93}
 \citeyear{riley93}) for all CESS sources but 
GLG046 and GLG152.  Additionally, most Group 1 LFVs have optical counterparts, 
not true for most of the CESS sources.  Sensitivity to lower $S_{325}$ does not
 necessarily help this explanation, since at lower $S_{325}$ normal galaxies 
dominate the radio population (Carole Jackson, private communication 2002) and 
these sources are not expect to vary with such magnitude.

If the CESS sources are LFVs, and the variation is not intrinsic, than the 
probable explanation for the low--frequency variability is interstellar 
scintillation (ISS).  ISS, caused by variations in electron density 
($\delta n_e$) along the line of sight, typically comes in two forms:
\begin{itemize}
\item Diffractive Interstellar Scintillation (DISS), which can cause flux 
modulations of up to 100\% over timescales on the order of minutes and over a 
bandwidth of the order of a MHz \citep{cordes98}.
\item RISS , the form of ISS believed 
to be behind most low--frequency variation (see Section \ref{variables} for 
more details), tends to cause a lower degree of flux modulation than DISS but 
it occurs over a much longer timescale (on the order of hours) and a much 
wider bandwidth (several MHz) than DISS \citep{cordes98}.
\end{itemize}
Neither form of ISS is a viable explanation for the CESS sources.  For the 
case of DISS, its limited bandwidth and short timescales make it an unlikely 
cause.  According to 
the NE2001 model for the electron density of the Milky Way, the bandwidth 
of DISS towards M31 is $\sim$0.4~kHz \citep{ne2001}.  Since the observations 
were done in spectral line mode with channel bandwidths of 10~kHz, the average 
effect of DISS over the entire bandwidth is negligible.  In addition, the 
short timescales of DISS argue against it as the cause of the CESS sources.  
Though extragalactic\footnote{In this context, ``extragalactic'' means outside 
the Milky Way, not outside both the Milky Way and M31.} sources have a 
DISS timescale roughly 3-30$\times$ larger than that of Galactic (Milky Way) 
objects \citep{cordes98}, increasing the DISS timescale from minutes to 
$\la$1~hour, since M31 was observed for a total of hour hours a 100\% increase 
in 325~flux density over this timescale would be insufficient to create the 
observed steep spectra of these sources.

RISS is unlikely to be the source of the CESS sources since the flux 
modulation typically produced by RISS is too small.  Flux density monitoring of
 pulsars have found that flux variation caused by RISS in distant pulsars is 
on the order of $\sim$5\%, insufficient to create the steep spectrum observed 
here.  Extragalactic LFVs that vary as a result of RISS typically have flux 
variations on the order of a few percent \citep[e.g.][]{rys1990}, again 
insufficient to create the steep spectra observed here.

While RISS and DISS are unlikely to be the cause of the CESS, this does not 
rule out ISS playing a role.  There have been a few instances, often called
 Extreme Scattering Events \citep{fiedler94a}, where a source increases 
in 
flux by more than a magnitude over a bandwidth of several MHz and over a 
timescale that ranges from minutes to hours.  The most extreme example of this 
is PSR B0655+64, whose 325~MHz flux density increased by a factor of $\sim$43 
for roughly an hour over a bandwidth of $\sim$4~MHz \citep{galama97}.  The 
reason for this sudden amplification is not known, but is believed to be the 
result of ISS rather than an intrinsic process because this pulsar has been 
observed to be stable over timescale of minutes to weeks \citep{galama97}.  If 
something similar occurred to a source during our observation, than the 
measured flux density would be on the order of $\sim$10$\times$ higher than the
 actual flux density.  To see if this is a possible explanation for the CESS 
sources, we calculated the spectral index needed to explain the measured 
higher--frequency flux densities assuming the observed 325~MHz flux densities 
is 10$\times$ higher than the intrinsic value.  The results are summarized in 
Table \ref{cessess}. For ten of the CESS sources, the required spectral index 
is reasonable ($\alpha \ga -0.7$).  While it is a possibility, such extreme 
events as that seen in PSR~B0655+64 are extremely rare -- the event described 
above was the only one seen in a long--term pulsar monitoring project 
\citep{galama97} --  so it is unlikely that something similar happened to many 
objects during our observation.  However, there is precedence for this sort of 
behavior in the direction of M31 -- in the 36W survey, it was observed that 
the flux density of 5C3.132 decreased by a factor of five in eight years 
\citep{36w}.

As a result of the analysis above, we do not believe that variability -- 
whether intrinsic or the result of ISS -- does a good job of explaining the 
observed steep spectrum for most of the CESS sources.  Therefore, we are left 
with Option \#2, that the CESS sources (or at least some of them) are 
intrinsically steep-spectrum sources.  The most common steep spectrum sources 
are radio pulsars, High--z Radio Galaxies (HzRGs), and Relic Radio galaxies 
(RRGs).

If a CESS source is a pulsar, then it must be unresolved and is in either M31 
or the Milky Way.  To test this possibility, we assume that a CESS source in 
the optical disk of M31 ($R_{M31} \leq 27$~kpc) is a pulsar in M31 -- meaning 
it has a distance of 780~kpc, while for a CESS source outside the optical disk 
of M31 we assumed it is a pulsar in the Milky Way -- implying a distance of 
$\sim$10 kpc.  To determine if the pulsar possibility is reasonable, we 
compared the luminosity of these sources with that of the brightest galactic 
pulsars -- at $\nu=400~\mbox{MHz}$ it is PSR B1302-64 with a beamed 
luminosity\footnote{Since radio emission from pulsars are not emitted 
isotropically, we used $L=S~\times~d^{2}$ not $L=S~\times~4\pi d^2$ to account 
for beaming.}  $L_{400}$ of 26100 mJy\ kpc$^2$, while at $\nu=1.4~\mbox{GHz}$ 
the brightest pulsar is PSR B0736-40 with a luminosity $L_{1400}$ of 9700 
mJy\ kpc$^2$ (numbers taken from the ATNF Pulsar Database).  The results of 
this analysis are shown in Table \ref{cesspsr}, whose columns are as follows:
\begin{description}
\item[Column 1:] GLG Name
\item[Column 2:] $R_{M31}$ in kpc, as defined in Section 4.1 in Paper I.
\item[Column 3:] $L_{400}$ in mJy kpc$^{2}$
\item[Column 4:] $L_{1400}$ in mJy kpc$^{2}$
\item[Column 5:] $\alpha_{L}^{400}$, the spectral index required for $L_{400}$
 of the source to be equal to that of PSR B1302-64.
\item[Column 6:] $\alpha_{L}^{1400}$, the spectral index required for 
$L_{1400}$ of the source to be equal to that of PSR B0736-40.
\end{description}
Both $L_{400}$ and $L_{1400}$ are the beamed luminosity.  The luminosities of 
CESS sources which might be pulsars in M31 are higher, by at least an order of 
magnitude, than that of the brightest observed Galactic pulsars.  The spectral 
index required to make these luminosities consistent with that of the 
brightest 400~MHz pulsar are much steeper than that of any known pulsar, those 
for some of these sources the spectral index required to make the luminosity 
consistent with that of the brightest 1.4~GHz pulsar is similar to that of the 
steepest spectrum radio pulsars \citep{lorimer95}.  For CESS sources outside 
the optical disk, their radio luminosities at a distance of 10~kpc is 
consistent with that of Milky Way pulsars in the Milky Way, but are unlikely 
to be pulsars because of their high Galactic latitude.  It is extremely 
premature to suggest that any of these sources may be pulsars, since deep 
optical imaging is needed to rule out the possibility that a source is a 
HzRG/relic radio galaxy, objects expected to be faint optical sources 
\citep{kaplan01}.

Another possibility is that the CESS sources are pulsars than emit giant 
pulses, which have been detected from a wide variety of pulsars -- from young 
pulsars like the Crab Pulsar \citep{lundgren95} to the millisecond pulsar 
PSR~B1937+21 \citep{kinkhabwala00}.  Even though giant pulses can be as much 
as $4000\times$ brighter than the average pulse, they do not significantly 
increase the integrated flux density \citep{lundgren95}.  The most extreme 
case of this is PSR B0540-69 in the LMC, which emits giant pulses but remains 
undetected in continuum 1400~MHz observations.  This pulsar has a spectral 
index $\alpha_{640}^{1380} \leq -4.4$ \citep{johnston03}, a similar limit as 
for many of the CESS sources in Table \ref{cesspsr}.

The second possibility for the intrinsic steep--spectrum explanation is that 
they are High-z Radio Galaxies (HzRGs), which were discussed in Section 
\ref{hzrgs}.  In that section, the steep--spectrum HzRG candidates discussed 
were all resolved radio sources with the morphology of an FR~II.  However, 
HzRG candidates often are much smaller than those sources -- the \citet{hzrg} 
survey of Ultra Steep Spectrum ($\alpha\leq-1.3$; USS) sources found that USS 
sources have a constant median angular size of $\sim$6\arcsec~between 
S$_{1400}$=10~mJy--1~Jy \citep{hzrg}.  Therefore, we expect to detect 
unresolved USS sources in this survey.  The density of CESS sources in this 
survey ($\sim$9500 sr$^{-1}$) is much higher than in other surveys, e.g. the 
density of USSs in the WENSS and NVSS survey is $\sim$151 sr$^{-1}$ 
\citep{hzrg}.  A higher density of steep--spectrum sources in the GLG survey is
 not completely unexpected since our survey is substantially deeper than the 
WENSS survey (a flux limit of $S_{325} \approx$3~mJy as opposed to $S_{325} 
\approx$18~mJy) and the 1.4 GHz surveys used for comparison are much 
deeper than the NVSS survey.  However, as mentioned earlier, starburst galaxies
 dominate the population of sources with $S_{1400} \la 10$~mJy 
-- which are not expected to have spectra as steep as these sources.

The third possibility for an intrinsically steep spectrum is relic radio 
galaxies.  These are objects which used to be FRII radio galaxies, but the 
central AGN turned off causing the relativistic electrons insides the lobes of 
these galaxies to lose energy through synchrotron emission and inverse Compton 
Scattering off the Cosmic Microwave Background (CMB).  The higher energy 
electrons lose energy faster than the lower energy ones, steepening the 
electron energy distribution and, consequently, the observed radio spectral 
index \citep{komis94}.  These sources can have spectral indices as steep as 
$-6 \la \alpha \la -5$, but tend to be rather large with 
$\theta_M \ga 10$\arcsec~\citep{kaplan00}.  Only GLG006, GLG022, GLG036, 
GLG059, and GLG305 have $\theta_M > 10$\arcsec, with GLG036 the largest with 
$\theta_M=16.6$\arcsec.  There is an R--band source on the edge of
GLG036, but it is unlikely the two are related.  Also, GLG036 is one
of the weakest CESS sources detected.

For the reasons stated above, we do not believe that is possible at this time 
to definitively state the nature of the observed CESS sources, and feel they 
are most likely a mix of the possibilities described above.  Deeper optical 
imaging and another 325~MHz radio observation of M31 would be very helpful in 
solving this mystery: optical imaging would determine whether or not these 
sources are galaxies (HzRGs tend to have $R \la 20$~mag), and another radio 
image would put a strong constraint on the variability of these sources.  
These sources are much fainter than previous steep--spectrum samples, and 
therefore are potentially extremely interesting.

\subsubsection{Anomalous Morphology and Multi--Wavelength Characteristics}
\label{weirdmorph}
In this section, we discuss sources with anomalous radio morphologies and/or 
multi-wavelength properties.

\paragraph{GLG015}
\label{glg015}
GLG015 is a complex source located on the edge of the optical disk of M31 
with a spectral index of $\alpha=-0.99\pm0.08$ and no spectral curvature, and 
has the morphology of a compact FRII.  However, GLG015 
is bordered on two sides by two clumps of H$\alpha$ and [SII] emission 
(see Figures \ref{halpgd} and \ref{siigd}) and is orientated such that it fits 
between them without overlap.  This emission has no counterparts in the R--band
 image of this region, or in any of the Far-IR, IR, and X--ray catalogs, but 
the bright spots in H$\alpha$/[SII] correspond to stars in the MagOpt catalog. 
 GLG015 most likely is a FRII coincidentally located in 
this position and orientation, if it is connected to the H$\alpha$/[SII] 
emission around it, the nature of this source is unknown.

\paragraph{GLG065}
\label{glg065}
GLG065 is a complex located just outside the optical disk of M31.  GLG065 has 
the most puzzling morphology of all sources in the GLG source list - as 
shown in Figure \ref{complex} it appears to be a collection of four, 
maybe five points sources arranged somewhat symmetrically.  It is unlikely that
 this source is a gravitational lens due to the the wide separation of 
the sources (at $\sim$30\arcsec~this would be one the most widely separated 
gravitational lenses known), and the lack of optical/IR emission expected from
 the lensing object.  It does have a passing resemblance to 3C315, a classic 
``X--shaped'' radio source \citep{leahy84}.  Another possible explanation is 
that GLG065 is a collection of SNRs in M31, although the lack of H$\alpha$ 
emission in this region makes this explanation unlikely.

\paragraph{GLG078}
\label{glg078}
GLG078 is a complex, steep spectrum ($\alpha=-1.57$) source with significant 
spectral turnover ($\varphi=0.10$) just beyond the optical disk of M31 and the 
morphology of an asymmetric FRII.  There is 
a 2MASS PSC source located next to the center of the radio emission - but there
 is no radio emission from the specific location of the IR source, and it has 
the IR colors of a star in the Milky Way.  The R--band image of GLG078 (shown 
in Figure \ref{redgd}) shows the IR source mentioned above as well as a source 
located at the SE edge of the radio emission.  We believe that GLG078 is an 
FR~II and that the R--band source is coincidental.  Higher-resolution radio 
images are required to better understand the nature of this source.

\section{Conclusion}
\label{conclusion}
In this paper we have presented the results of a four hour 325~MHz
survey of M31.  A statistical analysis of the 405 radio sources
detected in this survey, done in Paper I \citep{paper1}, revealed that
most of these sources are background radio galaxies.  In this paper we
present our attempt to determine the nature of these source using
their observed radio spectral properties, radio morphologies, and
their Far--IR/IR/Optical/X--ray properties of the detected sources.
With this information we classified 112 out of the 405 detected
sources, as summarized in Table \ref{srcsum}.

Sources in the GLG catalog could be placed into four broad categories: sources
 in M31, sources in the Milky Way, extragalactic sources, and anomalous 
sources.  In M31, we have identified five supernova remnant candidates and 
three pulsar wind nebula candidates, given in Section \ref{m31srcs}.  The 
supernova remnant candidates have properties (extent and radio luminosity) 
similar to that of known supernova remnants in the Milky Way, M33, LMC, and 
SMC as shown in Figure \ref{snrbmajflux}.  Two of the pulsar wind nebula 
candidates are more luminous than the Crab Nebula, but only by a factor of a 
few.  Further observations are needed to confirm the classification of these 
sources and determine the morphology and basic physical parameters.  The 
objects are the brightest of their class in M31, and as such are individually 
interesting.

In the Milky Way, we believed we have identified three radio stars based on 
optical and near-IR information and a Planetary Nebula based on radio 
morphology, as described in Section \ref{mwsrcs}.  A wide variety of 
extragalactic sources were identified, ranging from the ordinary (FRI and FRII 
radio galaxies, star-forming galaxies, and radio jets) to the exotic (a BL Lac
 candidate, a galaxy merger, a Giant Radio Galaxy candidate, and High-z Radio 
Galaxy Candidates, including Compact Steep Spectrum and Gigahertz Peaked 
Spectrum radio galaxies), using the criteria described in Section 
\ref{extragal}.  Last but not least, we have identified many anomalous sources 
 - ranging from highly variable sources (Section \ref{variables}) to compact 
($\theta_M \la 10$\arcsec) extremely steep spectrum ($\alpha \leq -1.6$) 
sources (Section \ref{cess}) to source with anomalous morphologies and/or 
multi-wavelength properties (Section \ref{weirdmorph}).

To summarize, in just a four-hour 325~MHz observation of M31 we have 
potentially identified the brightest supernova remnants and pulsar wind 
nebulae in these galaxy, as well as identified a class of steep spectrum 
sources whose nature is unknown.  In this paper, we have a detailed a technique
 for separating background radio sources from those intrinsic to M31, which 
will be extremely useful in future radio surveys of this galaxy.  While much 
remains unknown about the radio population of M31, this survey is an important 
step in understanding the properties of this galaxy.

\acknowledgements
 TJW Lazio acknowledges that basic research in radio astronomy at the NRL 
is supported by the Office of Naval Research.  The authors wish to
 thank the anonymous referee for his/her helpful comments, 
Andrew Hopkins for providing us with the latest version of $\mathcal{SFIND}$; 
Elias Brinks, Mike Garcia, Phil Kaaret, Albert Kong, Linda Schmidtobreick, 
Sergey Trudolyubov, Rene Walterbos, Ben Williams, C. Kevin Xu, and the SIMBAD 
help desk for providing us with source-lists and/or images; Carole Jackson for 
providing her model of the flux distribution of background radio galaxies at 
325~MHz; and Aaron Cohen, Elly Berkhuijsen, Jim Cordes, Rosanne DiStefano, 
Mike Garcia, Dan Harris, John Huchra, Namir Kassim, Albert Kong, Pat Slane, 
Krzysztof Stanek, Lorant Sjouwerman, Ben Williams, and Josh Winn for useful 
discussions; and Harvard CDF for computer access.

The National Radio Astronomy Observatory is a facility of the National Science 
Foundation operated under cooperative agreement by Associated Universities, 
Inc. 
This research has made use of NASA's Astrophysics Data System; of the SIMBAD 
database, operated at CDS, Strasbourg, France; of the NASA/IPAC Extragalactic 
Database (NED) which is operated by the Jet Propulsion Laboratory, California 
Institute of Technology, under contract with the National Aeronautics and 
Space Administration; of data products from the Two Micron All Sky Survey, 
which is a joint project of the University of Massachusetts and the Infrared 
Processing and Analysis Center/California Institute of Technology, funded by 
the National Aeronautics and Space Administration and the National Science 
Foundation; and of the NASA/ IPAC Infrared Science Archive, which is operated 
by the Jet Propulsion Laboratory, California Institute of Technology, under 
contract with the National Aeronautics and Space Administration.

\bibliographystyle{apj}

\clearpage
\begin{deluxetable}{ccc|cc}
\tablecolumns{5}
\tablewidth{0pc}
\tableheadfrac{0.05}
\tabletypesize{\scriptsize}
\tablecaption{Steep Spectrum ($\alpha<-1.3$)~GLG Sources  \label{steepspec}}
\tablehead{
\colhead{Name} & \colhead{Type} & \colhead{Flux [mJy]} & 
\colhead{$\alpha$} & \colhead{$\varphi$}}
\startdata
GLG002 &  U & 3.65$\pm$0.59 & -1.45$\pm$0.34 & 0.13$\pm$0.05 \\
GLG006 &  U & 18.12$\pm$0.58 & $\leq$-3.18 & \nodata \\
GLG007 &  U & 5.45$\pm$0.58 & $\leq$-2.37 & \nodata \\
GLG014 &  U & 23.42$\pm$0.69 & $\leq$-3.35 & \nodata \\
GLG022 & El & 11.62$\pm$1.79 & $\leq$-2.81 & \nodata \\
GLG031 &  C & 63.97$\pm$7.47 & -1.44$\pm$0.08 & \nodata \\
GLG036 & El & 38.38$\pm$2.59 & $\leq$-3.53 & \nodata \\
GLG047 &  U & 8.84$\pm$0.75 & -1.64$\pm$0.10 & \nodata \\
\enddata
\tablecomments{The spectral curvature parameter $\varphi$ was calculated only 
for those sources detected in three radio surveys, including this one.}
\end{deluxetable}

\begin{deluxetable}{ccc|cc}
\tablecolumns{5}
\tablewidth{0pc}
\tableheadfrac{0.05}
\tabletypesize{\scriptsize}
\tablecaption{Flat Spectrum ($-0.5<\alpha<0.5$)~GLG Sources  \label{flatspec}}
\tablehead{
\colhead{Name} & \colhead{Type} & \colhead{Flux [mJy]} & 
\colhead{$\alpha$} & \colhead{$\varphi$}}
\startdata
GLG011 &  U & 37.23$\pm$1.90 & -0.17$\pm$0.08 & -0.01$\pm$0.01 \\
GLG016 &  U & 3.87$\pm$0.50 & -0.07$\pm$0.08 & \nodata \\
GLG021 &  U & 8.57$\pm$0.76 & 0.16$\pm$0.48 & -0.07$\pm$0.07 \\
GLG027 &  U & 36.57$\pm$1.62 & -0.45$\pm$0.23 & -0.04$\pm$0.03 \\
GLG034 &  U & 12.95$\pm$0.46 & -0.29$\pm$0.04 & \nodata \\
GLG049 &  U & 7.99$\pm$0.66 & -0.00$\pm$0.14 & \nodata \\
GLG066 &  U & 4.25$\pm$0.49 & -0.49$\pm$0.17 & \nodata \\
GLG067 &  U & 14.29$\pm$0.85 & -0.29$\pm$0.10 & \nodata \\
\enddata
\tablecomments{The spectral curvature parameter $\varphi$ was calculated only 
for those sources detected in three radio surveys, including this one.}
\end{deluxetable}

\begin{deluxetable}{ccc|cc}
\tablecolumns{5}
\tablewidth{0pc}
\tableheadfrac{0.05}
\tabletypesize{\scriptsize}
\tablecaption{GLG Sources with Spectral Tunrnover ($\varphi \geq 
  3\sigma_{\varphi}$) \label{turnspec}}
\tablehead{
\colhead{Name} & \colhead{Type} & \colhead{Flux [mJy]} & 
\colhead{$\alpha$} & \colhead{$\varphi$}}
\startdata
GLG009 & ExG & 45.87$\pm$4.46 & -1.28$\pm$0.06 & 0.10$\pm$0.013 \\
GLG039 &  U & 25.98$\pm$1.12 & -1.21$\pm$0.10 & 0.11$\pm$0.016 \\
GLG046 &  U & 5.76$\pm$0.64 & -2.01$\pm$0.16 & 0.18$\pm$0.028 \\
GLG048 &  U & 16.00$\pm$0.72 & -1.18$\pm$0.14 & 0.08$\pm$0.019 \\
GLG050 &  C & 140.35$\pm$9.10 & -0.57$\pm$0.00 & -0.03$\pm$0.000 \\
GLG059 &  U & 30.69$\pm$0.94 & -1.77$\pm$0.08 & 0.17$\pm$0.011 \\
GLG077 &  U & 530.96$\pm$26.03 & -0.58$\pm$0.00 & -0.03$\pm$0.003 \\
GLG078 &  C & 109.30$\pm$8.91 & -1.57$\pm$0.00 & 0.10$\pm$0.000 \\
\enddata
\end{deluxetable}

\begin{deluxetable}{c|cc|cc|c}
\tablecolumns{6}
\tablewidth{0pc}
\tableheadfrac{0.05}
\tabletypesize{\scriptsize}
\tablecaption{Comparison with Near-IR, Far-IR, Optical and X--ray Surveys 
\label{mwcomp}}
\setlength{\tabcolsep}{0.02in}
\tablehead{
\colhead{Survey} & \colhead{\# of GLG sources} & 
\colhead{\# of GLG sources} & \colhead{\# of false} & 
\colhead{\# of false} & \colhead{Reference} \\
\colhead{} & \colhead{with counterparts} & 
\colhead{with good counterparts} & \colhead{counterparts} & 
\colhead{good counterparts}
}
\startdata
2MASS ESC & 21 & 13 & 12.00$\pm$3.74 & 7.00$\pm$1.60 & \cite{2MASSesc} \\
\nodata & 20 & 12 & 10.87$\pm$3.35 & 6.37$\pm$2.06 & \nodata \\
2MASS PSC & 44 & 29 & 33.87$\pm$3.13 & 20.12$\pm$2.64 & \cite{2MASSpsc} \\
\nodata & 17 & 17 & 1.50$\pm$1.60 & 1.50$\pm$1.60 & \nodata \\
\hline
``Knots'' & 33 & 12 & 39.87$\pm$4.94 & 10.37$\pm$3.20 & \cite{knots} \\
Xu \& Helou & 12 & 10 & 14.75$\pm$4.89 & 8.12$\pm$2.90 & \cite{xu} \\
IRAS-PSC & 5 & 5 & 1.00$\pm$1.19 & 1.00$\pm$1.19 & \cite{iraspsc} \\
IRAS-SSC & 2 & 2 & 0.87$\pm$0.99 & 0.75$\pm$1.03 & \cite{irasssc} \\
\enddata
\tablecomments{``RASS'' stands for {\it ROSAT} all sky survey, and {\it ROSAT} 
survey represents the Second {\it ROSAT} PSPC Survey of M31.  The other survey 
abbreviations are defined in the text.  The second line for the 2MASS ESC, 
2MASS PSC, RASS and {\it ROSAT} surveys are the results of this analysis only 
using ``U'' and ``El'' GLG sources.  The horizontal lines delineate different
wavebands.  The definition of a ``good'' counterpart is given in Section 
\ref{intro}.}
\end{deluxetable}

\begin{deluxetable}{cc||cc|ccc|cc}
\tablecolumns{9}
\tablewidth{0pc}
\tableheadfrac{0.05}
\tabletypesize{\scriptsize}
\tablecaption{IR Magnitudes and Colors of 2MASS ESC counterparts of GLG 
sources \label{jhkesc}}
\setlength{\tabcolsep}{0.02in}
\tablehead{
\colhead{GLG} & \colhead{Source} & \colhead{} & \colhead{} & 
\colhead{2MASS} & \colhead{ESC} & \colhead{Source} & 
\colhead{} \\ 
\colhead{Name} & \colhead{Type} & \colhead{Name} & \colhead{Offset
  [\arcsec]} & \colhead{J Mag.} & \colhead{H mag.} & \colhead{K Mag.}
& \colhead{J-H} & \colhead{H-K}} 
\startdata
GLG018 &  U & 2MASS-EXT00415185+4124424 & 0.45 & 14.9$\pm$0.19 & 14.1$\pm$0.22 
& 13.3$\pm$0.15 & 0.83 & 0.82 \\
GLG163 &  U & 2MASS-EXT00383316+4128509 & 0.23 & 13.1$\pm$0.09 &
12.2$\pm$0.11 & 11.7$\pm$0.10 & 0.85 & 0.55 \\
GLG184 &  U & 2MASS-EXT00380553+4120166 & 0.84 & 13.7$\pm$0.07 & 
12.8$\pm$0.08 & 12.4$\pm$0.09 & 0.97 & 0.33 \\
GLG203 &  U & 2MASS-EXT00415339+4021174 & 0.42 & 12.9$\pm$0.04 & 
12.1$\pm$0.05 & 11.6$\pm$0.05 & 0.79 & 0.52 \\
GLG204 & El & 2MASS-EXT00453674+4030033 & 0.30 & 15.4$\pm$0.11 & 
14.6$\pm$0.12 & 14.3$\pm$0.15 & 0.82 & 0.30 \\
GLG209 &  U & 2MASS-EXT00421611+4019360 & 0.48 & 14.3$\pm$0.08 & 
13.7$\pm$0.14 & 13.0$\pm$0.11 & 0.51 & 0.75 \\
\enddata
\end{deluxetable}

\begin{deluxetable}{cc||cc|ccccc|cc}
\tablecolumns{11}
\tablewidth{0pc}
\tableheadfrac{0.05}
\tabletypesize{\scriptsize}
\tablecaption{IR Magnitude and Colors of 2MASS PSC counterparts of GLG sources 
\label{jhkpsc}}
\setlength{\tabcolsep}{0.02in}
\tablehead{
\colhead{GLG} & \colhead{Source} & colhead{} & \colhead{} & 
\colhead{} & \colhead{2MASS} & \colhead{PSC} & \colhead{Sources} 
 & \colhead{} & \colhead{} & \colhead{} \\ 
\colhead{Name} & \colhead {Type} & \colhead{Name} & \colhead{Offset
  [\arcsec]} & \colhead{J Mag.} & \colhead{H Mag.} & 
\colhead{K Mag.} & \colhead{B Mag.} & \colhead{R Mag.} & 
\colhead{J-H} & \colhead{H-K}} 
\startdata
GLG002 &  U & 00422667+4118043 & 0.63 & 15.6$\pm$0.10 & 14.8$\pm$0.12
 & 14.2$\pm$0.10 & \nodata & \nodata & 0.81 & 0.60 \\
GLG003 &  U & 00423301+4111423 & 1.76 & 16.7$\pm$0.19 & 15.7$\pm$0.19
 & 15.2$\pm$0.20 & \nodata & \nodata & 1.00 & 0.43 \\
GLG017 &  C & 00420221+4126468 & 14.1 & 15.7$\pm$0.06 & 15.0$\pm$0.09
 & 14.7$\pm$0.10 & 18.9 & 18.5 & 0.74 & 0.24 \\
GLG018 &  U & 00415182+4124420 & 0.27 & 15.5$\pm$0.08 & 14.5$\pm$0.08
 & 13.8$\pm$0.06 & 17.6 & 19.1 & 1.00 & 0.68 \\
GLG030 &  C & 00413926+4130352 & 1.49 & 15.9$\pm$0.09 & 15.2$\pm$0.11
 & 14.5$\pm$0.08 & \nodata & \nodata & 0.68 & 0.68 \\
GLG050 &  C & 00450182+4124532 & 1.10 & 17.2$\pm$0.22 & 16.1$\pm$0.18
 & 15.8$\pm$0.21 & \nodata & \nodata & 1.03 & 0.38 \\
GLG078 &  C & 00453965+4112170 & 4.13 & 13.9$\pm$0.02 & 13.6$\pm$0.02
 & 13.6$\pm$0.03 & 15.6 & 14.5 & 0.26 & 0.04 \\
GLG079 &  U & 00425297+4149210 & 0.60 & 16.4$\pm$0.12 & 15.4$\pm$0.11
 & 14.7$\pm$0.10 & 19.5 & 18.5 & 0.93 & 0.69 \\
\enddata
\end{deluxetable}

\begin{deluxetable}{c|cc}
\tablecolumns{5}
\tablewidth{0pc}
\tableheadfrac{0.05}
\tablecaption{GLG sources with a counterpart in the WW~SNR catalog
  \label{optctrp}}
\tablehead{
\colhead{GLG} & \colhead{Optical} & \colhead{Counterpart} \\
\colhead{Name} & \colhead {Name} & \colhead{Offset [\arcsec]}}
\startdata
GLG011 & WW~29 & 52.8 \\
GLG020 & WW~47 & 2.21 \\
GLG037 & WW~12 & 1.83 \\
GLG068 & WW~10 & 2.25 \\
GLG123 & WW~60 & 41.4 \\
\enddata
\end{deluxetable}

\begin{deluxetable}{ccc||cc|cccc}
\tablecolumns{9}
\tablewidth{0pc}
\tableheadfrac{0.05}
\tabletypesize{\scriptsize}
\tablecaption{GLG sources with a Good Match in the MagOpt Catalog
\label{magstarctrp}}
\tablehead{
\colhead{} & \colhead{GLG Source} & \colhead{} & 
\colhead{} & \colhead{} & \colhead{MagOpt} & \colhead{Counterpart} & 
\colhead{} & \colhead{} \\ 
\colhead{Name} & \colhead {Type} & \colhead{$S_{325}$ [mJy]} & 
\colhead{Name} & \colhead{Offset [\arcsec]} & 
\colhead{B Mag.} & \colhead{V Mag.} & \colhead{R Mag.} & 
\colhead{I Mag.}}
\startdata
GLG008 &  U & 17.4 & MagOpt20716 & 3.20 & \nodata & \nodata & \nodata & 
19.6$\pm$0.12 \\
GLG014 &  U & 23.4 & MagOpt26706 & 2.15 & \nodata & \nodata & \nodata & 
20.1$\pm$0.19 \\
GLG019 &  U & 1223.6 & MagOpt31940 & 0.34 & \nodata & \nodata & \nodata
 & 20.2$\pm$0.13 \\
GLG023 &  U & 105.9 & MagOpt31494 & 1.09 & 14.9$\pm$0.04 & 14.0$\pm$0.02
 & 13.5$\pm$0.02 & 13.2$\pm$0.00 \\
GLG040 &  U & 10.2 & MagOpt13395 & 2.54 & 21.2$\pm$0.07 & 20.5$\pm$0.05
 & 20.0$\pm$0.11 & 19.5$\pm$0.10 \\
GLG061 & El & 23.2 & MagOpt12931 & 3.03 & \nodata & \nodata & \nodata & 
20.5$\pm$0.24 \\
GLG083 & El & 14.5 & MagOpt38317 & 3.04 & \nodata & \nodata & 
24.3$\pm$0.26 & 19.3$\pm$0.09 \\
GLG087 &  U & 14.7 & MagOpt18329 & 3.08 & \nodata & \nodata & \nodata & 
20.7$\pm$0.13 \\
\enddata
\tablecomments{The number assigned to each MagOpt source corresponds to its 
row number in Table 6 of \citet{magstars}.}
\end{deluxetable}

\begin{deluxetable}{cc||cc|cc|c}
\tablecolumns{7}
\tablewidth{0pc}
\tableheadfrac{0.05}
\tabletypesize{\scriptsize}
\tablecaption{Good X--ray counterparts of GLG sources \label{xraygdctrp}}
\setlength{\tabcolsep}{0.02in}
\tablehead{
\colhead{GLG} & \colhead{Source} & \colhead{} & \colhead{} &
\colhead{X--Ray} & \colhead{Sources} & \colhead{} \\ 
\colhead{Name} & \colhead{Type} & \colhead{Name} & \colhead{Offset
  [\arcsec]} & \colhead{HR1} & 
\colhead{HR2} & \colhead{$f_X$ [erg cm $^{-2}$ s$^{-1}$]} } 
\startdata
GLG005 &  U & r3-64 & 1.28 & 0.67$\pm$0.13 & 0.83$\pm$0.09 & \nodata \\
GLG011 &  U & r3-70 & 1.09 & 0.34$\pm$0.18 & 0.47$\pm$0.17 & \nodata \\
GLG020 &  U & XMM~North~40 & 1.13 & \nodata & \nodata & \nodata \\
GLG021 &  U & RXJ0041.7+4126 & 1.47 & \nodata & \nodata & 7.11$\times
10^{-14}$ \\
GLG023 &  U & XMM~North~46 & 0.49 & \nodata & \nodata & \nodata \\
GLG028 &  U & RXJ0041.6+4103 & 3.71 & \nodata & \nodata & 5.52$\times
10^{-14}$ \\
GLG041 &  U & RXJ0043.6+4054 & 10.9 & \nodata & \nodata & 1.22$\times
10^{-13}$ \\
GLG068 &  U & RXJ0040.7+4055 & 6.17 & \nodata & \nodata & $2.85\times
10^{-14}$\\
\enddata
\tablecomments{``RASS'' sources are from the {\it ROSAT} All--Sky survey 
(\cite{RASSfaint}; \cite{RASSbright}), ``RX'' sources are from the {\it ROSAT}
Second PSPC survey of M31 \citep{rosat}, ``r'' sources are from the 
\citet{acis} X--ray survey of M31, ``HRC'' sources are from the \citet{hrc} 
X--ray survey of M31, and ``XMM~North'' sources are from the {\it XMM} survey 
of M31 (Sergey Trudolyubov, private communication). $f_X$ is for the 0.1--2.0 
KeV band and assumes a power law with photon index $\Gamma=-2.0$ and 
$N_H=9\times10^{20} \mbox{cm}^{-2}$, and is calculated using data from the 
Second {\it ROSAT} PSPC Survey of M31 \citep{rosat}.}
\end{deluxetable}

\begin{deluxetable}{cc|cc||c|c|cc|c}
\tablecolumns{9}
\tablewidth{0pc}
\tableheadfrac{0.05}
\tabletypesize{\scriptsize}
\tablecaption{Radio Properties of M31 SNR and PWN Candidates \label{snrcan}}
\tablehead{
\colhead{Name} & \colhead{$R_{M31}$} & \colhead{$\alpha$} &  
\colhead{$\varphi$} & \colhead{Criterion} & \colhead{Size} & 
\colhead{$L_{325}$} & \colhead{$L_R$} & \colhead{EM} \\ 
\colhead{} & \colhead{[kpc]} & \colhead{} & \colhead{} &
\colhead{} & \colhead{[pc$\times$pc]} & \colhead{[mJy~kpc$^{2}$]} & 
\colhead{[${\mbox{ergs}}~\mbox{s}^{-1}$]} & \colhead{[pc~cm$^{-6}$]} 
}
\startdata
GLG020 & 3.5 & -0.83$\pm$0.05 & \nodata &  A~ & $<$23~$\times$~$<$20 & 
6.2$\times10^{7}$ & 2.3$\times10^{34}$ & \nodata \\
GLG037 & 9.4 & -0.95$\pm$0.09 & \nodata &  A~ & $<$37~$\times$~$<$26 & 
7.1$\times10^{7}$ & 2.1$\times10^{34}$ & \nodata \\
GLG056 & 12.2 & -0.60$\pm$0.14 & \nodata &  B~ & $<$21~$\times$~$<$18 & 
3.1$\times10^{7}$ & 2.3$\times10^{34}$ & \nodata \\
GLG123 & 13.0 & -0.76$\pm$0.17 & \nodata &  A~ & $<$20~$\times$~$<$17 & 
2.0$\times10^{8}$ & 9.0$\times10^{34}$ & \nodata \\
GLG193 & 13.6 & -0.93$\pm$0.14 & 0.07$\pm$0.02 &  D~ & $<$20~$\times$~$<$19 & 
6.6$\times10^{7}$ & 2.1$\times10^{34}$ & 2.1$\times10^{5}$ \\
\hline
GLG011 & 5.8 & -0.17$\pm$0.08 & \nodata &  HFV(2)~A~ & $<$20~$\times$~$<$18 & 
2.8$\times10^{8}$ & 1.2$\times10^{36}$ & \nodata \\
GLG068 & 8.4 & -0.25$\pm$0.09 & \nodata &  A~E~ & $<$21~$\times$~$<$19 & 
3.2$\times10^{7}$ & 9.7$\times10^{34}$ & \nodata \\
GLG198 & 15.4 & -0.08$\pm$0.20 & -0.07$\pm$0.02 &  C~ & $<$23~$\times$~$<$17
 & 9.0$\times10^{7}$ & 5.8$\times10^{35}$ & \nodata \\
\enddata
\tablecomments{The first five sources in this table are the SNR
  candidates, and the second three are the PWN candidates.  The
  ``Criterion'' column refers to the method of selecting the SNR/PWN
  candidates described in Section \ref{snrs}.  GLG011 is also a
  High-Frequency Variable (HFV) sources, described in Section \ref{variables}.}
\end{deluxetable}

\clearpage
\begin{deluxetable}{cccc||c|c||ccc}
\tablecolumns{9}
\tablewidth{0pc}
\tableheadfrac{0.05}
\tabletypesize{\scriptsize}
\tablecaption{Properties of FRI Candidates \label{frIcan}}
\tablehead{
\colhead{Name} & \colhead{$S_{325}$~[mJy]} & \colhead{$\alpha$} & 
\colhead{$\varphi$} & \colhead{Type} & \colhead{Criterion} & 
\colhead{IR?} & \colhead{Optical?} & \colhead{X--Ray?}}
\startdata
GLG060 & 82.9$\pm$6.03 & -0.78$\pm$0.11 & \nodata & FRI~ & A~ & N & N & N
 \\
GLG090 & 49.8$\pm$6.24 & -0.68$\pm$0.00 & -0.03$\pm$0.00 & FRI~ & A~ & N
 & N & Y \\
GLG122 & 18.9$\pm$12.37 & -1.25$\pm$0.15 & \nodata & FRI~ & A~ & Y & N & N
 \\
GLG186 & 710.2$\pm$35.47 & 0.98$\pm$0.03 & -0.14$\pm$0.00 & FRI~HzRG~ & 
 B~C~CSS & N & N & Y \\
GLG284 & 450.8$\pm$19.58 & -0.79$\pm$0.06 & \nodata & FRI~ &  B~ & N &
N & Y 
\enddata
\tablecomments{The criteria for determining if a source is a FRI radio
galaxy is described in Section \ref{agn}.  GLG186 is also a High-z
Radio Galaxy candidate (HzRG), as described in Section \ref{hzrgs}.}
\end{deluxetable}

\begin{deluxetable}{cccc||c|c||ccc}
\tablecolumns{9}
\tablewidth{0pc}
\tableheadfrac{0.05}
\tabletypesize{\scriptsize}
\tablecaption{Properties of FRII Candidates \label{frIIcan}}
\setlength{\tabcolsep}{0.02in}
\tablehead{
\colhead{Name} & \colhead{$S_{325}$~[mJy]} & \colhead{$\alpha$} & 
\colhead{$\varphi$} & \colhead{Type} & \colhead{Criterion} &
\colhead{IR?} & \colhead{Optical?} & \colhead{X--Ray?}
}
\startdata
GLG015 & 206.8$\pm$13.62 & -0.99$\pm$0.08 & \nodata &  FRII~Anom.~Source~ & 
 FRII~Morph.~Anom.~Source~ & N & N & N \\
GLG017 & 27.0$\pm$7.05 & -1.27$\pm$0.09 & \nodata &  FRII~NG~ & 
 FRII~Morph.~I~ & Y & N & N \\
GLG029 & 28.5$\pm$4.85 & -0.81$\pm$0.27 & \nodata &  FRII~ &  FRII~Morph.~
 & N & N & N \\
GLG030 & 76.3$\pm$8.81 & -0.54$\pm$0.10 & \nodata &  FRII~NG~ & 
 FRII~Morph.~I~ & Y & N & N \\
GLG050 & 140.3$\pm$9.10 & -0.57$\pm$0.00 & -0.03$\pm$0.00 &  FRII~NG~ & 
 FRII~Morph.~I~ & Y & N & N \\
GLG074 & 191.5$\pm$12.89 & -0.85$\pm$0.09 & \nodata &  FRII~ & 
 FRII~Morph.~ & N & N & Y \\
GLG078 & 109.3$\pm$8.91 & -1.57$\pm$0.00 & 0.10$\pm$0.00 & 
 FRII~HzRG~Anom.~Source~ &  FRII~Morph.~HzRG~Anom.~Source~ & Y & N & N \\
GLG085 & 247.5$\pm$15.69 & -0.99$\pm$0.09 & \nodata &  FRII~ & 
 FRII~Morph.~ & N & N & N \\
\enddata
\tablecomments{The criterion for determine if a source is a FRII radio
galaxy is described in Section \ref{agn}.  Several of these sources
fall in other categories as well, listed in the ``Type'' column.  In
this column, ``Anom. Source'' means the source has anomalous
morphological and/or multi-wavelength properties and is discussed in
Section \ref{weirdmorph}, ``NG'' means the source also meets the
criteria of a ``normal galaxy'' as defined in Section \ref{galaxies},
``HzRG'' means it is a High-z Radio Galaxy candidate as defined in
Section \ref{hzrgs}, ``LFV'' means it is a Low-Frequency Variable as
defined in Section \ref{variables}, and ``RT'' means it is part of a
Radio Triple.  The criteria for their inclusion in these categories
are also listed in the ``Criterion'' column, and are defined in their
respective sections.}
\end{deluxetable}

\begin{deluxetable}{cccc||c|c||cccc}
\tablecolumns{9}
\tablewidth{0pc}
\tableheadfrac{0.05}
\tabletypesize{\scriptsize}
\tablecaption{Properties of Normal Galaxies \label{rgcan}}
\setlength{\tabcolsep}{0.02in}
\tablehead{
\colhead{Name} & \colhead{$S_{325}$~[mJy]} & \colhead{$\alpha$} & 
\colhead{$\varphi$} & \colhead{Type} & \colhead{Criterion} & 
\colhead{IR?} & \colhead{Optical?} & \colhead{X--Ray?} & 
\colhead{EM~[pc cm$^{-6}$]}
}
\startdata
GLG002 & 3.65$\pm$0.59 & -1.45$\pm$0.34 & 0.13$\pm$0.05 &  NG~ &  I~ & Y
 & N & N & 4.0$\times10^{5}$ \\
GLG003 & 4.76$\pm$0.53 & -0.95$\pm$0.02 & \nodata &  NG~ &  I~ & Y & N & 
N & \nodata \\
GLG017 & 27.0$\pm$7.05 & -1.27$\pm$0.09 & \nodata &  FRII~NG~ & 
 FRII~Morph.~I~ & Y & N & N & \nodata \\
GLG018 & 10.6$\pm$0.76 & -1.02$\pm$0.22 & \nodata &  NG~NG~ &  I~II~ & Y
 & N & N & \nodata \\
GLG030 & 76.3$\pm$8.81 & -0.54$\pm$0.10 & \nodata &  FRII~NG~ & 
 FRII~Morph.~I~ & Y & N & N & \nodata \\
GLG041 & 191.1$\pm$7.32 & -0.97$\pm$0.08 & 0.03$\pm$0.01 &  NG~LFV~ & 
 III~1~ & N & N & Y & 1.0$\times10^{5}$ \\
GLG050 & 140.3$\pm$9.10 & -0.57$\pm$0.00 & -0.03$\pm$0.00 &  FRII~NG~ & 
 FRII~Morph.~I~ & Y & N & N & 1.0$\times10^{5}$ \\
GLG079 & 9.08$\pm$0.73 & -1.04$\pm$0.05 & \nodata &  NG~ &  I~ & Y & N & 
N & \nodata \\
\enddata
\tablecomments{The criteria used to determine if a source is a
  ``normal galaxy'' are defined in Section \ref{galaxies}.  Several of
these sources fall into other categories, as listed in the ``Type''
column.  FRII sources are discussed in Section \ref{agn}, ``LFV'' (Low
Frequency Variable) sources are discussed in Section \ref{variables},
as are ``HFV'' (High Frequency Variable) sources.  GLG105 has
previously been identified as a BL~Lac, discussed in Section
\ref{bllac}m and GLG247 has the morphology of a galaxy merger, as
discussed in Section \ref{glg247}.}
\end{deluxetable}

\begin{deluxetable}{cccc||c|c||ccc}
\tablecolumns{9}
\tablewidth{0pc}
\tableheadfrac{0.05}
\tabletypesize{\scriptsize}
\tablecaption{Properties of HzRG Candidates \label{hzrgcan}}
\setlength{\tabcolsep}{0.02in}
\tablehead{
\colhead{Name} & \colhead{$S_{325}$~[mJy]} & \colhead{$\alpha$} & 
\colhead{$\varphi$} & \colhead{Type} & \colhead{Criterion} & 
\colhead{IR?} & \colhead{Optical?} & \colhead{X--Ray?}
}
\startdata
GLG078 & 109.3$\pm$8.91 & -1.57$\pm$0.00 & 0.10$\pm$0.00 & 
 FRII~HzRG~Anom.~Source~ &  FRII~Morph.~USS~Anom.~Source~ & Y & N & N \\
GLG129 & 51.3$\pm$6.98 & -1.45$\pm$0.30 & -0.08$\pm$0.03 &  FRII~HzRG~ & 
 FRII~Morph.~USS~ & N & N & N \\
GLG187 & 73.6$\pm$8.33 & -1.62$\pm$0.09 & \nodata &  FRII~HzRG~LFV~ & 
 FRII~Morph.~USS~1~ & N & N & N \\
GLG296 & 203.1$\pm$13.47 & -1.60$\pm$0.12 & 0.04$\pm$0.02 &  FRII~HzRG~LFV~ & 
 FRII~Morph.~USS~1~ & N & N & N \\
\hline
GLG084 & 8.52$\pm$0.58 & -0.92$\pm$0.16 & 0.13$\pm$0.02 &  HzRG~ &  CSS/GPS~
 & N & N & N \\
GLG101 & 12.8$\pm$1.21 & -1.26$\pm$0.16 & 0.16$\pm$0.02 &  HzRG~ &  CSS/GPS~
 & N & N & N \\
GLG133 & 7.57$\pm$0.70 & -1.36$\pm$0.23 & 0.14$\pm$0.03 &  HzRG~ &  CSS/GPS~
 & N & N & N \\
GLG135 & 189.6$\pm$9.61 & -0.55$\pm$0.01 & 0.02$\pm$0.00 &  HzRG~ &  CSS/GPS~
 & N & N & N \\
\enddata
\tablecomments{The criteria for categorizing sources as High-z Radio
  Galaxy (HzRG) candidates are described in Section \ref{hzrgs}.
  Several of these sources fall under other classifications as well,
  as listed in the ``Type'' column.  FRI and FRII sources are
  described in Section \ref{agn}, Anomalous sources (``Anom. Source'')
are described in Section \ref{weirdmorph}, Low Frequency Variable
(``LFV'') sources are described in Section \ref{variables}, and
Compact Extremely Steep Spectrum (``CESS'') sources are described in
Section \ref{cess}.  The criteria for cataloging these sources as such
are listed under the ``Criterion'' column and described in their
respective sections.}
\end{deluxetable}

\begin{deluxetable}{ccc|cc||cc|cc||c}
\tablecolumns{10}
\tablewidth{0pc}
\tableheadfrac{0.05}
\tabletypesize{\scriptsize}
\tablecaption{Low-Frequency Variable sources \label{lfvcan}}
\setlength{\tabcolsep}{0.04in}
\tablehead{
\colhead{Name} & \colhead{Type} & \colhead{$R_{M31}$~[kpc]} & 
\colhead{$\alpha$} & \colhead{$\varphi$} & \colhead{WENSS Name} & 
\colhead{Offset~[\arcsec]} & \colhead{$S_{325}^{GLG}$~[mJy]} 
& \colhead{$S_{325}^{WENSS}$~[mJy]} & 
\colhead{$\frac{\Delta S_{325}}{\overline{S_{325}}}$}
}
\startdata
GLG009 & ExG & 5.4 & -1.28 & 0.10 &  WNB0040.8+4103 & 1.70 & 45.87$\pm$4.46 & 
84.00$\pm$4.00 & 0.58$\pm$0.09 \\
GLG041 & U & 21.9 & -0.97 & 0.03 &  WNB0040.9+4038 & 2.14 & 191.13$\pm$7.32 & 
254.00$\pm$3.90 & 0.28$\pm$0.03 \\
GLG187 & C & 52.8 & -1.62 & \nodata &  WNB0043.1+4020 & 10.97 & 73.69$\pm$8.33
 & 16.00$\pm$3.90 & 1.28$\pm$0.27 \\
GLG200 & U & 21.5 & -0.98 & 0.02 & \nodata & \nodata & 45.74$\pm$2.58 & 
$\leq$30 & $\geq$0.41 \\
GLG259 & U & 61.1 & -0.96 & \nodata & \nodata & \nodata & 43.98$\pm$2.14 & 
$\leq$30 & $\geq$0.37 \\
GLG280 & U & 66.6 & -0.70 & \nodata & \nodata & \nodata & 69.80$\pm$3.44 & 
$\leq$30 & $\geq$0.79 \\
GLG296 & C & 42.2 & -1.60 & 0.04 &  WNB0046.0+4120 & 0.73 & 203.17$\pm$13.47
 & 277.00$\pm$3.60 & 0.30$\pm$0.05 \\
GLG340 & U & 22.1 & -1.15 & 0.02 &  WNB0035.0+4008 & 2.80 & 263.31$\pm$11.66
 & 330.00$\pm$3.50 & 0.22$\pm$0.04 \\
\enddata
\tablecomments{The ``Type'' column refers to the 325~MHz morphology of
the sources, as described in Section \ref{intro}.  $R_{M31}$~is the
projected radius of the sources in M31, as defined in Section 4.1
Equation 3 of Paper I \citet{paper1}.}
\end{deluxetable}

\begin{deluxetable}{c||cc|ccc}
\tablecolumns{6}
\tablewidth{0pc}
\tableheadfrac{0.05}
\tabletypesize{\scriptsize}
\tablecaption{1.4 GHz Properties of Low-Frequency Variable sources
 \label{lfvcanhf}}
\setlength{\tabcolsep}{0.02in}
\tablehead{
\colhead{Name} & \colhead{1.4~GHz Counterparts} & 
\colhead{Offset~[\arcsec]} & \colhead{$S_{1400}$~[mJy]} & 
\colhead{$\overline{S_{1400}}$~[mJy]} & 
\colhead{$\frac{\Delta S_{1400}}{\overline{S_{1400}}}$}
}
\startdata
GLG009 & NVSS004336+412020,37W158C & 7.75,2.62 & 10.67$\pm$0.46,27.50$\pm$0.40
 & 19.08$\pm$0.43 & 0.88$\pm$0.03 \\
GLG041 & NVSS004341+405428,Braun~162,37W168 & 1.80,0.55,1.27 & 
66.11$\pm$0.46,54.00$\pm$2.70,66.50$\pm$3.40 & 65.77$\pm$0.45 & 0.19$\pm$0.06
 \\
GLG187 &  NVSS004552+403628 & 5.0 & 6.14$\pm$6.14 & \nodata & \nodata \\
GLG200 & NVSS004111+402410,37W075 & 0.55,1.12 & 13.47$\pm$0.46,15.50$\pm$0.90
 & 14.48$\pm$0.71 & 0.13$\pm$0.07 \\
GLG259 &  NVSS004550+402257 & 2.8 & 10.65$\pm$10.65 & \nodata & \nodata \\
GLG280 &  NVSS004624+402215 &  0.3 & 24.18$\pm$24.18 & \nodata & \nodata \\
GLG296 & NVSS004846+413715,37W249 & 1.44,0.82 & 30.45$\pm$0.46,21.90$\pm$2.60
 & 26.17$\pm$1.86 & 0.32$\pm$0.10 \\
GLG340 & NVSS003745+402515,37W008 & 1.45,2.25 & 61.11$\pm$0.46,56.40$\pm$11.30
 & 58.75$\pm$7.99 & 0.08$\pm$0.19 \\
GLG354 &  NVSS004948+41116 & 7.2
 & 3.05$\pm$3.05 & \nodata & \nodata \\
GLG358 &  NVSS003544+413015 &  2.7 & 21.67$\pm$21.67 & \nodata & \nodata \\
GLG374 &  NVSS003522+41165 & 0.5 & 25.73$\pm$25.73 & \nodata & \nodata \\
GLG393 &  NVSS003453+413131 &  0.8 & 99.77$\pm$99.77 & \nodata & \nodata \\
GLG398 &  NVSS005043+41282 &  0.4 & 27.83$\pm$27.83 & \nodata & \nodata \\
\enddata
\end{deluxetable}

\begin{deluxetable}{c|cc||ccc||c}
\tablecolumns{7}
\tablewidth{0pc}
\tableheadfrac{0.05}
\tabletypesize{\scriptsize}
\tablecaption{High-Frequency Variable sources \label{hfvcan}}
\setlength{\tabcolsep}{0.02in}
\tablehead{
\colhead{Name} & \colhead{$\alpha$} & \colhead{$\varphi$} & 
\colhead{1.4~GHz Counterparts} & \colhead{Offset~[\arcsec]} &
\colhead{$S_{1400}$~[mJy]} & \colhead{$\frac{\Delta
    S_{1400}}{\overline{S_{1400}}}$}
}
\startdata
GLG004 & -0.84 & 0.04 & NVSS004318+411729,37W158A & 7.96,11.73 & 
3.76$\pm$0.46,12.60$\pm$0.30 & 1.07$\pm$0.08 \\
GLG009 & -1.28 & 0.10 & NVSS004336+412020,37W158C & 7.75,2.62 & 
10.67$\pm$0.46,27.50$\pm$0.40 & 0.88$\pm$0.03 \\
GLG011 & -0.17 & -0.01 & NVSS004251+412633,Braun~103,37W144 & 1.25,0.31,4.87
 & 20.11$\pm$0.46,25.21$\pm$0.24,23.70$\pm$1.00 & 0.21$\pm$0.02 \\
GLG019 & -0.71 & -0.01 & NVSS004218+412926,Braun~61,37W115 & 0.22,0.18,0.49 & 
363.30$\pm$0.46,351.46$\pm$1.45,373.40$\pm$15.00 & 0.06$\pm$0.04 \\
GLG105 & -0.74 & 0.00 & NVSS004013+40505,37W051 & 5.02,6.46 & 
39.52$\pm$0.46,51.80$\pm$1.70 & 0.26$\pm$0.03 \\
GLG266 & -2.25 & \nodata & NVSS004126+401235,37W083B & 4.32,1.47 & 
2.72$\pm$0.46,41.10$\pm$3.10 & 1.75$\pm$0.22 \\
\hline
GLG161 & -0.58 & -0.02 & NVSS003932+40441,37W036 & 2.45,3.44 & 
9.71$\pm$0.46,9.70$\pm$0.60 & 0.00$\pm$0.07 \\
\enddata
\end{deluxetable}

\begin{deluxetable}{ccc||cc|cc|cc}
\tablecolumns{9}
\tablewidth{0pc}
\tableheadfrac{0.05}
\tabletypesize{\scriptsize}
\tablecaption{325~MHz Properties of High-Frequency Variable sources 
\label{hfvcanlf}}
\setlength{\tabcolsep}{0.02in}
\tablehead{
\colhead{GLG Name} & \colhead{Type} & \colhead{$R_{M31}$} & 
\colhead{WENSS Name} & \colhead{Offset~[\arcsec]} & 
\colhead{$S_{325}^{GLG}$~[mJy]} & \colhead{$S_{325}^{WENSS}$~[mJy]} & 
\colhead{$\overline{S_{325}}$~[mJy]} & 
\colhead{$\frac{\Delta S_{325}}{\overline{S_{325}}}$}
}
\startdata
GLG004 & U & 4.2 & \nodata & \nodata & 22.85$\pm$0.70 & \nodata & \nodata & 
\nodata \\
GLG009 & Ex & 5.4 &  WNB0040.8+4103 & 1.70 & 45.87$\pm$4.46 & 84.00$\pm$4.00
 & 64.93$\pm$4.23 & 0.58$\pm$0.09 \\
GLG011 & U & 5.8 &  WNB0040.1+4110 & 4.38 & 37.23$\pm$1.90 & 22.00$\pm$3.90 & 
29.61$\pm$3.06 & 0.51$\pm$0.15 \\
GLG019 & U & 12.3 &  WNB0039.5+4113 & 0.35 & 1223.64$\pm$61.20 & 
1190.00$\pm$3.90 & 1206.82$\pm$43.36 & 0.02$\pm$0.05 \\
GLG105 & C & 11.0 & \nodata & \nodata & 115.49$\pm$8.11 & \nodata & \nodata & 
\nodata \\
GLG266 & U & 30.5 & \nodata & \nodata & 96.05$\pm$2.13 & \nodata & \nodata & 
\nodata \\
\enddata
\tablecomments{The ``Type'' column refers to the 325~MHz morphology of
the sources, as described in Section \ref{intro}.  $R_{M31}$~is the
projected radius of the sources in M31, as defined in Section 4.1
Equation 3 of Paper I \citet{paper1}.}
\end{deluxetable}

\begin{deluxetable}{ccc|ccc||c|c||ccc}
\tablecolumns{11}
\tablewidth{0pc}
\tableheadfrac{0.05}
\tabletypesize{\scriptsize}
\tablecaption{Properties of CESS Sources \label{cesscan}}
\setlength{\tabcolsep}{0.02in}
\tablehead{
\colhead{Name} & \colhead{Type} & \colhead{$R_{M31}$~[kpc]} & 
\colhead{$S_{325}$~[mJy]} & \colhead{$\alpha$} & 
\colhead{$\varphi$} & \colhead{Source Type} & \colhead{Criteria} & 
\colhead{IR?} & \colhead{Optical?} & \colhead{X--Ray?}
}
\startdata
GLG006 &  U & 8.5 & 18.1$\pm$0.58 & $\leq$-3.18 & \nodata &  CESS~ &  CESS~ & 
N & N & N \\
GLG007 &  U & 9.0 & 5.45$\pm$0.58 & $\leq$-2.37 & \nodata &  CESS~ &  CESS~ & 
N & N & N \\
GLG014 &  U & 7.3 & 23.4$\pm$0.69 & $\leq$-3.35 & \nodata &  CESS~ &
CESS~ & N & N & N \\
GLG022 & El & 16.3 & 11.6$\pm$1.79 & $\leq$-2.81 & \nodata &  CESS~ &  CESS~ & 
N & N & N \\
GLG036 & El & 20.8 & 38.3$\pm$2.59 & $\leq$-3.53 & \nodata &  CESS~ & 
 CESS~ & N & Y & N \\
GLG046 &  U & 18.8 & 5.76$\pm$0.64 & -2.01$\pm$0.16 & 0.18$\pm$0.02 &  CESS~ & 
 CESS~ & N & N & N \\
GLG047 &  U & 25.1 & 8.84$\pm$0.75 & -1.64$\pm$0.10 & \nodata &  CESS~ &  
CESS~ & N & N & N \\
GLG055 &  U & 27.4 & 18.7$\pm$1.10 & $\leq$-2.01 & \nodata &  CESS~ &  CESS~ & 
 N & N & N \\
\enddata
\tablecomments{The ``Type'' column refers to the 325~MHz morphology of
the sources, as described in Section \ref{intro}.  $R_{M31}$~is the
projected radius of the sources in M31, as defined in Section 4.1
Equation 3 of Paper I \citet{paper1}.  The criteria for classifying a
sources as a CESS sources is described in Section \ref{cess}.
GLG059 is also part of a radio jet, as described in Section
\ref{agn}, while GLG152 is also a High-z Radio Galaxy (HzRG)
candidate, as described in Section \ref{hzrgs}.  The criteria for
these identifications are listed in the ``Criteria'' column, and
discussed in the respective sections.}
\end{deluxetable}

\begin{deluxetable}{cc||ccc|cc}
\tablecolumns{7}
\tablewidth{0pc}
\tableheadfrac{0.05}
\tabletypesize{\scriptsize}
\tablecaption{High Frequency Variability explanation for CESS Sources
  \label{cesshfv}}
\tablehead{
\colhead{Name} & \colhead{$S_{325}$~[mJy]} & 
\colhead{$S_{1400}^{exp}$~[mJy]} & 
\colhead{$S_{1400}^{meas}$~[mJy]} & 
\colhead{Survey} & 
\colhead{$\frac{\Delta S_{1400}}{\overline{S_{1400}}}$} & 
\colhead{$\overline{\mbox{Var}_{1400}}$} 
}
\startdata
GLG006 & 18.1 & 8.73 & $\leq$0.15 & Braun & $\geq$1.93 & $\geq$78 \\
GLG007 & 5.45 & 2.62 & $\leq$0.15 & Braun & $\geq$1.78 & $\geq$22 \\
GLG014 & 23.4 & 11.2 & $\leq$0.15 & Braun & $\geq$1.94 & $\geq$101 \\
GLG022 & 11.6 & 5.59 & $\leq$0.16 & Braun & $\geq$1.88 & $\geq$49 \\
GLG036 & 38.3 & 18.4 & $\leq$0.18 & Braun & $\geq$1.96 & $\geq$166 \\
GLG046 & 5.76 & 2.77 &       1.72 & Braun &       0.46 &       9 \\
GLG047 & 8.84 & 4.26 &      0.74 & Braun &       1.40 &       31 \\
GLG055 & 18.7 & 9.02 & $\leq$1.01 & Braun & $\geq$1.59 & $\geq$72 \\
\enddata
\end{deluxetable}

\begin{deluxetable}{cc||ccc|cc}
\tablecolumns{7}
\tablewidth{0pc}
\tableheadfrac{0.05}
\tabletypesize{\scriptsize}
\tablecaption{Low Frequency Variability explanation for CESS Sources
  \label{cesslfv}}
\tablehead{
\colhead{Name} & \colhead{$S_{325}$~[mJy]} & 
\colhead{$S_{1400}^{meas}$~[mJy]} & \colhead{Survey} & 
\colhead{$S_{325}^{exp}$~[mJy]} & 
\colhead{$\frac{\Delta S_{325}}{\overline{S_{325}}}$} & 
\colhead{$\overline{\mbox{Var}_{325}}$} 
}
\startdata
GLG006 & 18.1 & $\leq$0.15 & Braun & 0.31 & $\geq$1.93 & $\geq$161 \\
GLG007 & 5.45 & $\leq$0.15 & Braun & 0.31 & $\geq$1.78 & $\geq$46 \\
GLG014 & 23.4 & $\leq$0.15 & Braun & 0.31 & $\geq$1.94 & $\geq$210 \\
GLG022 & 11.6 & $\leq$0.16 & Braun & 0.34 & $\geq$1.88 & $\geq$102 \\
GLG036 & 38.3 & $\leq$0.18 & Braun & 0.38 & $\geq$1.96 & $\geq$345 \\
GLG046 & 5.76 & 1.72 & Braun & 3.58 & 0.46 & 19 \\
GLG047 & 8.84 & 0.74 & Braun & 1.54 & 1.40 & 66 \\
GLG055 & 18.7 & $\leq$1.01 & Braun & 2.10 & $\geq$1.59 & $\geq$151 \\
GLG059 & 30.6 & 12.8 & NVSS & 26.6 & 0.14 & 50 \\
GLG062 & 24.9 & $\leq$0.29 & Braun & 0.61 & $\geq$1.90 & $\geq$221 \\
GLG104 & 16.4 & $\leq$1.34 & 37W & 2.78 & $\geq$1.42 & $\geq$56 \\
GLG140 & 52.4 & $\leq$2.50 & NVSS & 5.18 & $\geq$1.64 & $\geq$591 \\
GLG149 & 7.31 & $\leq$0.25 & Braun & 0.53 & $\geq$1.72 & $\geq$61 \\
GLG152 & 7.68 & 4.04 & NVSS & 8.40 & 0.08 & 9 \\
GLG160 & 13.7 & $\leq$0.15 & Braun & 0.31 & $\geq$1.91 & $\geq$122 \\
GLG171 & 18.8 & $\leq$0.23 & Braun & 0.49 & $\geq$1.89 & $\geq$167 \\
GLG263 & 26.2 & $\leq$2.50 & NVSS & 5.18 & $\geq$1.34 & $\geq$263 \\
GLG289 & 11.0 & $\leq$2.50 & NVSS & 5.18 & $\geq$0.72 & $\geq$73 \\
GLG294 & 11.5 & $\leq$2.50 & NVSS & 5.18 & $\geq$0.75 & $\geq$79 \\
GLG305 & 26.9 & $\leq$2.50 & NVSS & 5.18 & $\geq$1.35 & $\geq$272 \\
GLG319 & 10.2 & $\leq$0.25 & Braun & 0.52 & $\geq$1.80 & $\geq$87 \\
GLG405 & 22.7 & $\leq$2.50 & NVSS & 5.18 & $\geq$1.25 & $\geq$219 \\
\enddata
\end{deluxetable}

\begin{deluxetable}{cc||ccc}
\tablecolumns{5}
\tablewidth{0pc}
\tableheadfrac{0.05}
\tabletypesize{\scriptsize}
\tablecaption{Extreme Scattering Event explanation for CESS Sources
  \label{cessess}}
\tablehead{
\colhead{Name} & \colhead{$S_{325}$~[mJy]} & 
\colhead{$S_{325}^{ESE}$~[mJy]} & 
\colhead{$S_{1400}^{meas}$~[mJy]} & \colhead{$\alpha^{ESE}$}
}
\startdata
GLG006 & 18.1 & 1.81 & $\leq$0.15 & $\leq$-1.70 \\
GLG007 & 5.45 & 0.54 & $\leq$0.15 & $\leq$-0.87 \\
GLG014 & 23.4 & 2.34 & $\leq$0.15 & $\leq$-1.88 \\
GLG022 & 11.6 & 1.16 & $\leq$0.16 & $\leq$-1.32 \\
GLG036 & 38.3 & 3.83 & $\leq$0.18 & $\leq$-2.07 \\
GLG046 & 5.76 & 0.57 & 1.72 & 0.75 \\
GLG047 & 8.84 & 0.88 & 0.74 & -0.11 \\
GLG055 & 18.7 & 1.87 & $\leq$1.01 & $\leq$-0.41 \\
GLG059 & 30.6 & 3.06 & 12.8 & 0.98 \\
GLG062 & 24.9 & 2.49 & $\leq$0.29 & $\leq$-1.46 \\
GLG104 & 16.4 & 1.64 & $\leq$1.34 & $\leq$-0.13 \\
GLG140 & 52.4 & 5.24 & $\leq$2.50 & $\leq$-0.50 \\
GLG149 & 7.31 & 0.73 & $\leq$0.25 & $\leq$-0.72 \\
GLG152 & 7.68 & 0.76 & 4.04 & 1.13 \\
GLG160 & 13.7 & 1.37 & $\leq$0.15 & $\leq$-1.51 \\
GLG171 & 18.8 & 1.88 & $\leq$0.23 & $\leq$-1.42 \\
GLG263 & 26.2 & 2.62 & $\leq$2.50 & $\leq$-0.03 \\
GLG289 & 11.0 & 1.10 & $\leq$2.50 & $\leq$0.55 \\
GLG294 & 11.5 & 1.15 & $\leq$2.50 & $\leq$0.53 \\
GLG305 & 26.9 & 2.69 & $\leq$2.50 & $\leq$-0.05 \\
GLG319 & 10.2 & 1.02 & $\leq$0.25 & $\leq$-0.95 \\
GLG405 & 22.7 & 2.27 & $\leq$2.50 & $\leq$0.06 \\
\enddata
\end{deluxetable}

\begin{deluxetable}{cc||cc|cc}
\tablecolumns{6}
\tablewidth{0pc}
\tableheadfrac{0.05}
\tabletypesize{\scriptsize}
\tablecaption{Pulsar explanation for CESS Sources \label{cesspsr}}
\tablehead{
\colhead{Name} & \colhead{$R_{M31}$~[kpc]} & 
\colhead{$L_{400}$~[mJy~kpc$^2$]} & 
\colhead{$L_{1400}$~[mJy~kpc$^2$]} & $\alpha_L^{400}$ & 
\colhead{$\alpha_L^{1400}$}
}
\startdata
GLG006 & 8.5 & $\leq$5.6$\times10^{6}$ & $\leq$1.0$\times10^{5}$ & 
$\leq$-29.1 & $\leq$-4.8 \\
GLG007 & 9.0 & $\leq$2.0$\times10^{6}$ & $\leq$1.0$\times10^{5}$ & 
$\leq$-23.3 & $\leq$-3.9 \\
GLG014 & 7.3 & $\leq$7.1$\times10^{6}$ & $\leq$1.0$\times10^{5}$ & 
$\leq$-30.3 & $\leq$-4.9 \\
GLG022 & 16.3 & $\leq$3.9$\times10^{6}$ & $\leq$1.1$\times10^{5}$ & 
$\leq$-26.9 & $\leq$-4.5 \\
GLG036 & 20.8 & $\leq$1.1$\times10^{7}$ & $\leq$1.3$\times10^{5}$ & 
$\leq$-32.7 & $\leq$-5.3 \\
GLG046 & 18.8 & 2.3$\times10^{6}$ & 1.8$\times10^{5}$ & -23.5 &  -4.0 \\
GLG047 & 25.1 & 3.8$\times10^{6}$ & 4.9$\times10^{5}$ & -25.6 & -4.3 \\
GLG059 & 26.0 & 1.2$\times10^{7}$ & 1.3$\times10^{6}$ & -31.6 & -5.1 \\
\enddata
\tablecomments{The sources above the horizontal line lie within the
  optical disk of M31, while the sources below the line don't.}
\end{deluxetable}

\begin{table}
\caption{ Classification Summary \label{srcsum}}
\begin{center}
\begin{tabular}{rrcc}
\hline
\hline
{\bf Number of Classified Sources:} & & {\bf   112 } & \\
\hline
{\it M31 Sources}: & & {\it 8 } & \\
 & SNR Candidates: & & 5 \\
 & PWN Candidates: & & 3 \\
\hline
{\it Milky~Way Sources}: & & {\it 4 } & \\
 & Radio Stars: & & 3 \\
 & PN Candidates: & & 1 \\
\hline
{\it Extragalactic Sources}: & & {\it 66 } & \\
 & FRIs: & & 6 \\
 & FRIIs: & & 24 \\
 & Radio Jets: & & 5 (11) \\
 & Radio Triples: & & 1 (2) \\
 & Normal Galaxies: & & 23 \\
 & HzRG Candidates: & & 16 \\
 & Galaxy Mergers: & & 1 \\
 & BL Lacs: & & 1 \\
 & GRG Candidates: & & 1 (2) \\
\hline
{\it Variable Sources:} & & {\it 15} & \\
 & Low Frequency Variables: & & 10 \\
 & High Frequency Variables: & & 6 \\
\hline
{\it Anomalous Sources:} & & {\it 25} & \\
 & CESS sources: & & 22 \\
 & Anom. Morph./Multiwave. Props.: & & 3 \\
\end{tabular}
\end{center}
{\scriptsize {\it Note:} Some GLG sources fall into multiple categories. SNR 
stands for Supernova Remnant, PWN stands for Pulsar Wind Nebula, PN
stands for Planetary Nebula, HzRG stands for High--z Radio Galaxy, and
CESS stands for ``Compact Extremely Steep Spectrum.''  The number is
parenthesis indicate the number of GLG sources that fall into a
certain category, e.g. five radio jets were detected which comprise 11
GLG sources since, in some case, the components of the jets were
classified as separate GLG sources.} \\
\end{table}


\clearpage
\newpage
\figcaption{The location of steep--spectrum sources (upper left), 
flat--spectrum sources (upper right), and sources with spectral turnover 
(bottom), as defined in Section \ref{specclass}.  The names are to the right 
of the source location, and the optical image of M31 is from the Palomar All 
Sky Survey. As mentioned in Section \ref{specclass}, the clustering of
sources with spectral turnover around the optical disk of M31 is a
selection effect, further described in the text. \label{specsrcs}}

\figcaption[]{Greyscale images of all Complex GLG sources overlaid with 
contours.  Ellipses denote the size and orientation of the source as 
determined by the {\sc miriad} task {\sc sfind}.  All the GLG sources are 
roughly the same 
size because {\sc sfind} determined the properties using the smoothed 
maps described in Section 2.2.2 in Paper I.  The intensity scale is in Jy 
beam$^{-1}$.  The values of the contours change for each source, and 
correspond to $\frac{A}{i}$, where $A$ is the peak~flux of the source, and $i$ 
is a counter that begins at $i=1$ and increases by 1 for sources with 
$\frac{I}{\sigma_{RMS}}\leq10$, by 2 for sources with 
$10<\frac{I}{\sigma_{RMS}}\leq20$, and by 4 for sources with 
$\frac{I}{\sigma_{RMS}}>20$, until $\frac{A}{i}<\sigma_{RMS}$.\label{complex}}

\figcaption{The IR colors of all ESC sources within 1$^\circ$ of M31 
({\it top--left}), of good PSC counterparts of GLG sources ({\it top--right}),
of PSC sources within 5\arcmin~of M31 ({\it bottom--left}), and of PSC sources 
$>$90\arcmin~away from M31 ({\it bottom--right}).  In the top--left graph, 
diamonds are good ESC counterparts of GLG sources and crosses are good PSC 
counterparts of GLG sources, and in the bottom graphs crosses indicate good 
counterparts of GLG sources. \label{ircolors}}

\figcaption[]{R--band (Greyscale) image of GLG sources with optical 
counterparts, overlaid with radio contours (only for Complex sources) and 
ellipses that show the extent and orientation of the GLG source.  The colorbar 
is in ADU counts.  Since the R--band images were not calibrated, the 
conversion between ADU counts and flux is unknown. (Ben Williams, private 
communication) The contours change for each source, and correspond to 
$\frac{A}{2i}$, where $A$ is the peak~flux of the source, and $i$ is a counter 
that begins at $i=1$ and increases by 0.5 for sources with 
$\frac{I}{\sigma_{RMS}}\leq10$, by 2 for sources with 
$\frac{I}{\sigma_{RMS}}\leq20$, by 3 for sources with 
$20<\frac{I}{\sigma_{RMS}}\leq50$, and by 10 for sources with 
$50<\frac{I}{\sigma_{RMS}}$, until $\frac{A}{i}<\sigma_{RMS}$. \label{redgd}}

\figcaption[]{H$\alpha$ image of GLG sources with an H$\alpha$ counterpart.  
The colorbar is in ADU counts, and the conversion to flux is 
$3.28\times10^{-14} \frac{\rm erg}{\rm cm^2~ADU~s}$ (Ben Williams, private 
communication).  The contours change for each source, but follow the formula 
given in Figure \ref{redgd}. \label{halpgd}}

\figcaption[]{[SII] image of GLG sources with a [SII] counterpart.  The 
colorbar is in ADU counts, and the conversion to flux is 
$2.89\times10^{-14} \frac{\rm erg}{\rm cm^2~ADU~s}$ (Ben Williams, private 
communication).  The contours change for each source, but follow the formula 
given in Figure \ref{redgd}. \label{siigd}}

\figcaption[]{Radio contour images of all the Complex sources with
  good X--ray counterparts.    The cross represents a {\it
  Chandra}~HRC sources, while the ellipses represent {\it ROSAT}~PSPC
  sources.  The contour values change for each source, and correspond
  to $\frac{A}{i}$, where $A$ is the peak~flux of the source, and $i$
  is a counter that begins at $i=1$ and increases by 1 for sources
  with $\frac{I}{\sigma_{RMS}}\leq10$ until
  $\frac{A}{i}<\frac{\sigma_{RMS}}{2}$, and by 2 for sources with
  $10<\frac{I}{\sigma_{RMS}}\leq30$, and by 4 for sources with
  $\frac{I}{\sigma_{RMS}}>30$, until
  $\frac{A}{i}<\sigma_{RMS}$.\label{xraycnt}}

\figcaption{325~MHz Greyscale images of SNR (top two rows) and PWN
  (bottom row) candidates in the GLG source list.  The intensity scale
  is in Janskys.  The ellipse corresponds to the orientation and
  extent of the GLG source. \label{snrpics}}

\figcaption{The position of M31 SNR and PWN candidates with respect 
to the optical disk of M31.  Optical image courtesy of the Palomar All--Sky 
Survey. \label{snrpos}}

\figcaption{$S_{325}$ vs $\theta_M$ of Milky Way (Xs), M33 
(diamonds), LMC (triangles), and SMC (squares) supernova remnants if placed at 
the distance of M31.  SNR candidates in the GLG sources are denoted by the 
thick ``+'' signs.  Arrows indicate the upper limit on $\theta_M$ for GLG SNR 
candidates.  The line represents the observational limits of the GLG survey -- 
only SNRs with a flux density $S_{325} > 3~$mJy beam$^{-1}$ will be detected.
\label{snrbmajflux}}

\figcaption{325~MHz Radio image ({\it left}) and J--Band image ({\it right}) of
 GLG347.  The black contour indicates the location and extent of GLG347 and the
 grey contours correspond to $S_{325}=1, 2, 3, 4~\mbox{and}~5$ \mjyb.
\label{glg347pic}}

\figcaption{325~MHz grey scale of Radio Triple GLG220 and GLG212.  The black 
ellipses indicate the location and extent of the GLG sources, and the white 
contours correspond to $S_{325}=2, 3, 4, 5, 7$ \mjyb. \label{radtrip}}

\figcaption{Radio Jets GLG031/GLG033 ({\it upper--left}, previous page),  
GLG045/GLG051({\it upper--right}, previous page), GLG054/GLG059 
({\it bottom--left}, previous page), GLG266/GLG269 ({\it bottom--right}, 
previous page), GLG270/GLG271/GLG275 (this page).  The black ellipses 
are the GLG sources plus Braun 3 (middle ellipse in GLG031/GLG033), and the 
white contours are the same as in Figure \ref{radtrip}. \label{jetspic}}

\figcaption{Radio ({\it left}) and Palomar All Sky Survey optical image 
({\it right}) of GLG247 overlaid with radio contours (white) whose levels 
are $S_{325}=2, 3, 4, 5, 7$ \mjyb.  The black ellipse corresponds to the size 
and orientation of the GLG source.\label{glg247pic}}

\figcaption{Radio image of BL~Lac candidate GLG105 ({\it upper--left}), 
H$\alpha$ image of GLG105 ({\it upper--right}; courtesy of Ben Williams), and 
broadband SED of GLG105 ({\it bottom}). The black contours in the top two 
images are $S_{325}=2.5, 5, ..., 15$ \mjyb.  The arrows in the bottom image 
are upper--limits, while the diamond and triangle are the expected value for 
an FRI and FRII, respectively, using the procedure described in Section 
\ref{ir}. \label{glg105pic}}

\figcaption{\scriptsize NVSS image ({\it top--right}), Palomar All-Sky Survey 
image 
({\it top--left}), and 325~MHz radio image of Giant Radio Galaxy candidate 
GLG242/GLG260 ({\it bottom}).  White contours are $S_{325}=2.5, 5, ..., 15$ 
\mjyb.\label{grgpic}}

\figcaption{Raw ({\it top--row}) and Binned ({\it bottom--row}) Distribution of
 $\frac{\Delta S}{\overline{S}}$ vs. $R_{M31}$ at 325~MHz ({\it left}) and 
1400~MHz ({\it right}).  \label{varpics}}

\figcaption{325~MHz images of CESS sources. The ellipses correspond to the 
size and orientation of the GLG source. \label{cesspics}}

\end{document}